\documentclass[a4paper, 12pt]{article}

\usepackage[left=25mm, right=25mm, top=30mm, bottom=30mm]{geometry}
\usepackage[utf8]{inputenc}
\usepackage[english]{babel}
\usepackage{amsmath}
\usepackage{amsfonts}
\usepackage{dsfont}
\usepackage{graphicx}
\usepackage{subcaption}
\usepackage{cite}
\usepackage[colorlinks, allcolors=blue, linktocpage=true]{hyperref}

\numberwithin{equation}{section}

\renewcommand{\O}{\mathcal{O}}
\newcommand{\Otilde}{\widetilde{\mathcal{O}}}
\newcommand{\Pitilde}{\widetilde{\Pi}}
\newcommand{\jbar}{{\bar{\jmath}}}
\renewcommand{\hbar}{{\bar{h}}}
\newcommand{\wbar}{{\bar{w}}}
\newcommand{\zbar}{{\bar{z}}}
\DeclareMathOperator{\im}{im}
\DeclareMathOperator*{\disc}{disc}

\makeatletter
\newsavebox{\@brx}
\newcommand{\llangle}[1][]{\savebox{\@brx}{\(\m@th{#1\langle}\)}%
  \mathopen{\copy\@brx\kern-0.5\wd\@brx\usebox{\@brx}}}
\newcommand{\rrangle}[1][]{\savebox{\@brx}{\(\m@th{#1\rangle}\)}%
  \mathclose{\copy\@brx\kern-0.5\wd\@brx\usebox{\@brx}}}
\makeatother

\title{The momentum-space conformal bootstrap in 2d}
\author{\\ Marc Gillioz}
\date{\vspace{5mm} \emph{School of Engineering, University of Applied Sciences and Arts \\
of Western Switzerland HES-SO, 1950 Sion, Switzerland}}

\begin{document}

\maketitle

\vspace{5mm}

\begin{abstract}
    A new bootstrap equation in 2-dimensional conformal field theory is derived starting from the momentum-space representation of the correlation functions. Since Wightman functions are not crossing-symmetric, the analyticity properties of the commutator are leveraged instead to obtain a relation between two distinct operator product expansions. The procedure requires evaluating a 4-point function with two light-like momenta. The result is an asymmetric equation valid for arbitrary theories in 1d and 2d. The new crossing equation admits two simple projections onto orthogonal bases of Jacobi polynomials, reproducing known sets of analytic functionals. One of these, with zeros at double-twist dimensions, was known only as a contour integral. The closed-form expression found in this work is new. We provide a few examples of applications of the new crossing equation.
\end{abstract}

\thispagestyle{empty}

\newpage
\tableofcontents

\section{Introduction}

Among all quantum field theories, the ones with conformal symmetry stand out for many reasons. They are found at the end point of renormalization group flows, describe continuous phase transitions, and are dual to gravity in anti de-Sitter space. But maybe most importantly, they can be rigorously defined in terms of an infinite algebra of operators characterized by a scaling dimension and spin quantum numbers. Remarkably, requiring the closure of this algebra is sometimes sufficient for solving an entire conformal field theory --- this goes by the name of conformal bootstrap~\cite{Belavin:1984vu, Rattazzi:2008pe}.

The development of the conformal bootstrap program has been rapid in recent years and some of its results are truly remarkable~\cite{Poland:2022qrs, Chang:2024whx}. Nevertheless, our analytical understanding of the mechanisms behind the conformal bootstrap is still incomplete. It is therefore crucial to develop it further~\cite{Hartman:2022zik}.

One direction that has been investigated in recent years is the momentum-space representation of the conformal correlation functions. It was initially noticed that the Fourier transform of Euclidean correlation functions admits interesting representations, both in closed form for 3-point functions \cite{Coriano:2013jba, Bzowski:2013sza} and in integral form for higher-point functions~\cite{Bzowski:2020kfw}. However, the Fourier integral interferes with the nice convergence properties of the operator product expansion (OPE) in Euclidean space, so these representations have not proven useful in the context of the conformal bootstrap. The situation is different in Minkowski space-time, where Wightman functions do admit an OPE both in the position- and momentum-space representations~\cite{Gillioz:2019lgs, Gillioz:2020wgw}. Alas, Wightman functions lack another crucial property required by the bootstrap approach: the permutation symmetry of the correlation functions, synonym with associativity of the operator algebra. For this reason, existing results in Minkowski momentum space rely on the delicate connection between Wightman functions and time-ordered correlators, such as the optical theorem~\cite{Gillioz:2016jnn, Gillioz:2018kwh, Meltzer:2020qbr}, dispersion formulae~\cite{Meltzer:2021bmb}, or LSZ reduction~\cite{Gillioz:2020mdd}, but a true conformal bootstrap in momentum space has never been achieved so far.

The goal of this work is to propose such a momentum-space bootstrap, working with Wightman functions and replacing the missing permutation symmetry with a property coming from the analyticity of the commutator, a direct consequence of the causality axiom of quantum field theory.
The present work focuses on theories with one time and one space directions only. Even though the method introduced below can be directly applied to conformal field theory in higher dimensions, many technicalities are simpler in 2d. Of course, 2d is also the domain in which the traditional bootstrap excels, thanks to powerful techniques involving Virasoro symmetry and its generalizations, or the modular invariance of the torus partition function. Our approach does not rely on such techniques in an essential way, though: it only assumes global conformal invariance in flat space-time, even though the integer spacing of conformal weights is used in some examples given below. 

\subsection*{Summary of results}

The calculations in this work all involve correlation functions of local operators characterized by pairs of conformal weight $(h, \hbar)$ that can in principle be distinct for each operator. However, for the sake of clarity, here we highlight the main results with identical scalar operators.

The idea is very similar to the ordinary conformal bootstrap. Starting from the 4-point function of a scalar operator $\phi$ with conformal weights $(h_\phi, h_\phi)$, we obtain two distinct expansions in conformal blocks that depend on the conformal weights $(h, \hbar)$ of the exchanged operator $\mathcal{O}$, multiplied by the square of the OPE coefficients $C_{\phi\phi\mathcal{O}}$.
A specificity of 2-dimensional conformal field theory is that the correlation functions, and hence the conformal blocks, can be factorized into holomorphic (left-moving) and anti-holomorphic (right-moving) parts that each depend on $h$ or $\hbar$ only.
We obtain the equation
\begin{equation}
    \sum_\O C_{\phi\phi\mathcal{O}}^2 S_h(w) S_{\hbar}(\bar{w})
    = \sum_\O C_{\phi\phi\mathcal{O}}^2 T_h(w) T_{\hbar}(\bar{w}),
    \qquad
    w, \bar{w} \in (0, 1).
    \label{eq:2dcrossing:identical}
\end{equation}
The holomorphic blocks $S_h$ and $T_h$ are functions of a single kinematic variable $w$ restricted to the interval $(0, 1)$. As we shall see, $w$ is the dimensionless ratio between the (left-moving) momentum of one of the external operators and the center-of-mass energy of the $2 \to 2$ scattering process. Similarly, the anti-holomorphic blocks $S_{\hbar}$ and $T_{\hbar}$ depend on $\bar{w}$, defined in terms of right-moving momenta.

The s-channel conformal block $S_h$ is a hypergeometric function multiplied by a power of $w$ and a ratio of gamma functions,
\begin{equation}
    S_h(w) = \frac{\Gamma(2h)}{\Gamma(h)^2 \Gamma(2h_\phi)} w^{2h_\phi - 1}
    {}_2F_1(1 - h, h; 2 h_\phi; w).
    \label{eq:S:identical}
\end{equation}
It differs from the t-channel block,
\begin{equation}
    T_h(w) = \frac{1}{\Gamma(2h_\phi - h) \Gamma(h)}
    w^{h - 1} {}_2F_1( h - 2h_\phi + 1, h; 2 h; w).
    \label{eq:T:identical}
\end{equation}
The derivation of equations~\eqref{eq:2dcrossing:identical}, \eqref{eq:S:identical}, and \eqref{eq:T:identical} is the subject of Section~\ref{sec:crossingequation} below.

The two functions $S_h$ and $T_h$ are reminiscent of the position-space blocks~\cite{Dolan:2000ut, Dolan:2003hv} and, in fact, a non-trivial connection with them can be established through the analytic functionals of Ref.~\cite{Mazac:2016qev}. As a reminder, the usual conformal blocks are also hypergeometric functions multiplying non-integer powers of the cross-ratio $z$. They can be analytically continued to the whole complex plane in $z$ except for two branch cuts at $z \in (-\infty, 0)$ and $z \in (1, \infty)$. We show in Section~\ref{sec:Euclideanconnection} that eqs.~\eqref{eq:S:identical} and \eqref{eq:T:identical} correspond to a particular integral transform of the discontinuity along either of these branch cuts.
With hindsight, it is actually rather simple to recover eq.~\eqref{eq:2dcrossing:identical} from the position-space crossing equation once the transform is given. However, it is probable that the transform would have never been found without the momentum-space origin of the equation.

The first consequence of this remarkable connection with the Euclidean branch cuts is that the t-channel block \eqref{eq:T:identical} has zeros at $h = 2h_\phi + n$ with integer $n \in \mathbb{N}$, as visible from the first gamma function in its denominator. This corresponds to the exchange of double-twist operators of the form $\phi \partial^n \bar{\partial}^{\bar{n}} \phi$.
The second, unfortunate consequence is that the absolute convergence of the Euclidean conformal block expansion is lost in eq.~\eqref{eq:2dcrossing:identical}. While the t-channel expansion is absolutely convergent --- eq.~\eqref{eq:T:identical} actually has similar asymptotics as the Euclidean blocks when $h \to \infty$ --- the s-channel expansion is only convergent in a distributional sense: it does not converge point-wise at fixed $w$, but only after integrating against a suitable functions in the domain $w, \bar{w} \in (0,1)$.

Interestingly, there exists a natural basis of functions that give a nice reformulation of the crossing equation~\eqref{eq:2dcrossing:identical}: after integrating against specific Jacobi polynomials in $w$, we get a family of equations
\begin{equation}
    \sum_\O C_{\phi\phi\mathcal{O}}^2 S_{h, j} S_{\hbar, \jbar}
    = \sum_\O C_{\phi\phi\mathcal{O}}^2 T_{h, j} T_{\hbar, \jbar},
    \qquad
    \forall~j, \jbar = 0, 1, 2, \ldots
    \label{eq:2dcrossing:j:identical}
\end{equation}
In this case, the integrated s-channel block takes the simple form
\begin{equation}
    S_{h, j} = \frac{\Gamma(2h)}{\Gamma(h)^2 \Gamma(2h_\phi - h) \Gamma(h + 2h_\phi - 1)}
    \frac{1}{(2h_\phi - h + j)(2h_\phi + h - 1 + j)}.
    \label{eq:Sj:identical}
\end{equation}
It has zeros at every $h = 2h_\phi + n$ with $n \in \mathbb{N}$, except when $n = j$, where it is finite.
The t-channel block is not as simple, but a closed-form expression can be given in terms of the generalized hypergeometric function:
\begin{equation}
    \begin{split}
        T_{h, j} &= \frac{\Gamma(h - 2h_\phi + 1)}
        {\Gamma(2h_\phi - h) \Gamma(h - 2h_\phi + 1 - j) \Gamma(h + 2h_\phi + j) } \\
        & \qquad\times
        {}_4F_3\left( \begin{array}{c}
            h, h, h - 2h_\phi + 1, h + 2h_\phi - 1 \\
            2h, h - 2h_\phi + 1 - j, h + 2h_\phi + j
        \end{array}; 1 \right).
    \end{split}
    \label{eq:Tj:identical}
\end{equation}
The zeros at $h = 2h_\phi + n$ with $n \in \mathbb{N}$ are also there. As a consequence, the crossing equation takes a particularly simple form in (generalized) free field theories: of all the composite operators entering the OPE $\phi \times \phi$, only a single operator of the schematic form $\phi \partial^j \bar{\partial}^{\jbar} \phi$ contributes to eq.~\eqref{eq:2dcrossing:j:identical}. It must therefore cancel against the identity contribution, given by the limit $h \to 0_+$,
\begin{equation}
    S_{0, j} = 0,
    \qquad\qquad
    T_{0, j} = \frac{(-1)^j}{\Gamma(2h_\phi)^2}.
    \label{eq:T0j:identical}
\end{equation}
This leads to an immediate and elegant solution to crossing in generalized free field theory, a feature that otherwise requires using the $AdS_{d+1}$ dual of the theory~\cite{Fitzpatrick:2011dm}
or performing an analytic continuation of the Euclidean conformal blocks in the light-cone limit~\cite{Fitzpatrick:2012yx, Komargodski:2012ek, Pal:2022vqc, vanRees:2024xkb}.

This family of conformal crossing equations was actually discovered in Ref.~\cite{Mazac:2019shk}, where it was given as a contour integral in cross-ratio space. To the best of our knowledge, the closed-form expressions \eqref{eq:Sj:identical} and \eqref{eq:Tj:identical} are new.
They form a very useful basis for constructing various functionals that can help bootstrap interacting theories.
We show that the known free fermion optimal functionals can be recovered numerically from this basis, and derive a new family of functionals that computes the OPE coefficients of primary operators in the vacuum module. The results agree with Virasoro algebra computations, even though only global conformal invariance is used in the construction of the functional.
They are presented in Section~\ref{sec:newbasis}, together with an explanation of the origin of eqs.~\eqref{eq:2dcrossing:j:identical}, \eqref{eq:Sj:identical}, \eqref{eq:Tj:identical}.

Our conclusions are given in Section~\ref{sec:conclusions}, where we emphasize that the momentum-space approach to crossing in conformal field theory will be even more powerful in dimensions $d > 2$. Finally, a few technical appendices contain calculations that did not fit in the main text.

\section{The momentum-space crossing equation}
\label{sec:crossingequation}

We begin with the derivation of equations \eqref{eq:2dcrossing:identical}, \eqref{eq:S:identical}, and \eqref{eq:T:identical}. This derivation relies on first principles of quantum field theory, namely causality expressed as the commutation of local operators at space-like separation, and the fact that correlation functions are tempered distributions that can be Fourier transformed between the position and momentum representations~\cite{Kravchuk:2020scc, Kravchuk:2021kwe}. Conformal symmetry provides the additional ingredients: an operator product expansion (OPE) that is absolutely convergent in position space, and the ability to expand 4-point functions into conformal blocks that cover the contributions of all descendants of a given primary operator.

%%%%%%%%%%%%%%%%%%%%%%%%%%%%%%%%%%%%%%%%%%%%%%%%%%%

\subsection{Analyticity from causality}

Let us consider the commutator of two local bosonic operators
\begin{equation}
    \big[ \mathcal{O}_a(x_a), \mathcal{O}_b(x_b) \big],
\end{equation}
and its Fourier transform
\begin{equation}
    \big[ \Otilde_a(p_a), \Otilde_b(p_b) \big]
    = \int d^2x_a d^2x_b e^{-i (p_a \cdot x_a + p_b \cdot x_b)} \big[ \mathcal{O}_a(x_a), \mathcal{O}_b(x_b) \big],
    \label{eq:commutator:momentum}
\end{equation}
where $p \cdot x = p^0 x^0 + p^1 x^1$.
The principle of causality in quantum field theory implies that the commutator vanishes at space-like separation. If we denote $x = x_a + x_b$ and $y = x_b - x_a$, this means
\begin{equation}
    \big[ \mathcal{O}_a\left(\tfrac{x-y}{2} \right),
    \mathcal{O}_b\left(\tfrac{x+y}{2} \right) \big]
    = 0
    \quad \text{for} \quad |y^0| < | y^1 |.
    \label{eq:causality}
\end{equation}
Note that the same principle applies to fermionic operators with a minus sign, replacing the commutator with an anti-commutator.

In momentum space, denoting $p_a = p - q$ and $p_b = p + q$, so that $p$ and $q$ are the momenta respectively conjugate to $x$ and $y$, we have
\begin{equation}
    \big[ \Otilde_a(p - q), \Otilde_b(p + q) \big] 
    = \int d^2x d^2y e^{-i (p \cdot x + q \cdot y)}
    \big[ \mathcal{O}_a\left(\tfrac{x-y}{2} \right),
    \mathcal{O}_b\left(\tfrac{x+y}{2} \right) \big].
\end{equation}
Since the integrand vanishes over part of the domain in $y$, this commutator has special analyticity properties in $q$. This is an operator statement, meaning that it is true when the commutator is part of any correlation function.
In conformal field theory, higher-point correlation functions of primary operators can be reduced to lower-point ones by means of the operator product expansion (OPE), and therefore all the consequences of causality are contained in the 4-point function
\begin{equation}
    \langle 0 | \Otilde_1(p_1) \big[ \Otilde_2(p - q), \Otilde_3(p + q) \big] 
    \Otilde_4(p_4) | 0 \rangle.
    \label{eq:4ptcommutator}
\end{equation}
The most general analyticity properties of this object in the variable $q$ are encoded in an integral representation derived by Jost and Lehmann~\cite{Jost:1957yis}.%
\footnote{Jost and Lehmann consider operators with a gap in their energy spectrum, but the results can be applied to the gapless case as well. Dyson obtained an even more general representation in the case of operators with distinct mass gaps~\cite{Dyson:1958uwa}.}
This representation takes a simple form if the states surrounding the commutator are momentum eigenstates, as in eq.~\eqref{eq:4ptcommutator}, but it should be interpreted as a tempered distribution in $p_1$ and $p_4$ that can be Fourier-transformed back to position space if needed.

\subsubsection*{Retarded commutator}
Instead of dealing with the Jost-Lehmann representation directly, it is often more convenient to work with the \emph{retarded} commutator defined with the help of the step function $\Theta$,
\begin{equation}
    \Theta(-y^0) \big[ \mathcal{O}_a\left(\tfrac{x-y}{2} \right),
    \mathcal{O}_b\left(\tfrac{x+y}{2} \right) \big] = 0
    \quad \text{for} \quad y^0 \geq -| y^1 |.
    \label{eq:retardedcommutator}
\end{equation}
As indicated, the retarded commutator vanishes everywhere outside the past light cone in $y$. Its Fourier transform is denoted
\begin{equation}
    \big[ \Otilde_a(p - q), \Otilde_b(p + q) \big]_R
    \equiv \int d^2x d^2y \, e^{-i (p \cdot x + q \cdot y)}
    \Theta(-y^0) \big[ \mathcal{O}_a\left(\tfrac{x-y}{2} \right),
    \mathcal{O}_b\left(\tfrac{x+y}{2} \right) \big]
\end{equation}
By eq.~\eqref{eq:retardedcommutator}, the Fourier transform of the retarded commutator must be analytic in $q$ over the forward tube, that is in the complex domain defined by
\begin{equation}
    \im q^0 > | \im q^1 |.
\end{equation}
In this domain, the Fourier integral is namely damped by the exponential $\exp(y \cdot \im q)$, and, provided that the retarded commutator is part of a correlation function that is a tempered distribution, it defines an analytic function of $q$.%

In momentum space, the retarded commutator is related to the regular commutator through a convolution in $q^0$,
\begin{equation}
    \big[ \Otilde_a(p - q), \Otilde_b(p + q) \big]_R
    = \frac{1}{2\pi i} \int d^2p' d^2q'
    \frac{1}
    {q'^0 - q^0}
    \delta^2(p' - p) \delta(q'^1 - q^1)
    \big[ \Otilde_a(p' - q'), \Otilde_b(p' + q') \big].
\end{equation}
The convolution integral is well-defined over the domain of analyticity with $\im q^0 > 0$.
Since it defines an analytic \emph{function} of $q$ in the forward tube, not just a \emph{distribution}, it can be evaluated at a particular point in $q = (q^0, q^1)$.
Choosing without loss of generality $q^0 = i$, the causality condition \eqref{eq:causality} is finally equivalent to the following condition:
\begin{equation}
    \int dq^0 \frac{1}{q^0 - i}
    \big[ \Otilde_a(p - q), \Otilde_b(p + q) \big]
    \qquad \text{is analytic in}~q^1~\text{in the domain}~| \im q^1 | < 1.
    \label{eq:analyticity}
\end{equation}

\subsubsection*{Sketch of the crossing equation}

The idea behind the momentum-space crossing equation is to evaluate \eqref{eq:analyticity} inside a 4-point function of conformal primary operators, separate the commutator into two Wightman functions
\begin{equation}
    \langle 0 | \Otilde_1(p_1) \Otilde_2(p - q)
    \Otilde_3(p + q) \Otilde_4(p_4) | 0 \rangle
    -
    \langle 0 | \Otilde_1(p_1) \Otilde_3(p + q)
    \Otilde_2(p - q) \Otilde_4(p_4) | 0 \rangle,
    \label{eq:4ptfunctions}
\end{equation}
and expand each of them in conformal blocks using the OPE.
The fact that the integral over $q^0$ converges for each individual piece is not obvious:  $(q^0 - i)^{-1}$ is not a Schwartz test function, so there are no first-principle quantum field theory arguments for convergence. But in CFT we can examine each conformal block individually and verify that it does indeed converge if a certain condition on the operators' scaling dimension is fulfilled, given in eq.~\eqref{eq:convergencecondition} below.
We also observe that each conformal block is discontinuous at a particular real value of $q^1$, denoted $q^1_*$, and that the discontinuity is preserved by the infinite sum over conformal blocks. Such a discontinuity is in contradiction with the analyticity condition \eqref{eq:analyticity}, so it must precisely cancel between the two Wightman functions.
In other words, we obtain an equation of the form,
\begin{equation}
    \begin{split}
        &\disc_{q^1 = q^1_*} \int dq^0 \frac{1}{q^0 - i}
        \langle 0 | \Otilde_1(p_1) \Otilde_2(p - q)
        \Otilde_3(p + q) \Otilde_4(p_4) | 0 \rangle \\
        & \qquad = 
        \disc_{q^1 = q^1_*} \int dq^0 \frac{1}{q^0 - i}
        \langle 0 | \Otilde_1(p_1) \Otilde_3(p + q)
        \Otilde_2(p - q) \Otilde_4(p_4) | 0 \rangle.
        \label{eq:crossing:disc}
    \end{split}
\end{equation}
This is a crossing equation in the bootstrap sense: it relates two correlation functions in which the order of two operators are exchanged. Each side can be expanded into an infinite sum of unknown OPE coefficients multiplying functions that are completely determined by conformal symmetry in terms of the conformal weights of the operators involved.

%%%%%%%%%%%%%%%%%%%%%%%%%%%%%%%%%%%%%%%%%%%%%%%%%%%

\subsection{Example: the Gaussian 4-point function}
\label{sec:Gaussianexample}

The existence of the discontinuity leading to the crossing equation \eqref{eq:crossing:disc} is best understood with a simple example. Consider a 4-point function involving two distinct local operators $\O_a$ and $\O_b$, and assume that it factorizes into 2-point functions as in a Gaussian theory,
\begin{equation}
    \langle 0 | \O_a(x_1) \O_a(x_2) \O_b(x_3) \O_b(x_4) | 0 \rangle
    \equiv \langle 0 | \O_a(x_1) \O_a(x_2) | 0 \rangle \langle 0 | \O_b(x_3) \O_b(x_4) | 0 \rangle.
    \label{eq:4pt:Gaussian}
\end{equation}
If $\O_a$ and $\O_b$ are primary operators in a conformal field theory, the 2-point functions are fixed by symmetry up to a conventional normalization:
\begin{equation}
    \langle 0 | \O(x) \O(y) | 0 \rangle
    = \Pi_h(x^+ - y^+) \Pi_\hbar(x^- - y^-),
    \label{eq:2pt:Pi}
\end{equation}
where we have introduced light-cone coordinates $x^\pm = x^0 \pm x^1$ and defined, following the conventions of Ref.~\cite{Gillioz:2019iye},
\begin{equation}
    \Pi_\alpha(x) = \lim_{\varepsilon \to 0_+}
    \frac{e^{-i \pi \alpha}}{(x - i \varepsilon)^{2\alpha}}.
    \label{eq:Pi:x}
\end{equation}
$h$ and  $\hbar$ are the conformal weights of the operator $\O$, related to its scaling dimension $\Delta$ and spin $\ell$ by $h = \frac{1}{2} (\Delta + \ell)$ and $\hbar = \frac{1}{2} (\Delta - \ell)$. The phase in the definition of $\Pi_\alpha$ is chosen such that the operators are real: the 2-point function obeys $\langle 0 | \O(x) \O(y) | 0 \rangle^* = \langle 0 | \O(y) \O(x) | 0 \rangle$.
With this choice, the correlation function is real at space-like separation if the spin is integer ($\ell = h - \hbar \in \mathbb{Z}$) and purely imaginary if it is half-integer, which is consistent with the bosonic and fermionic commutation relations \eqref{eq:causality}.

The position-space 2-point function is a tempered distribution for any $h, \hbar > 0$: it gives finite results when integrated against test functions in $x$ and $y$. This means that the Fourier transform exists and is also a tempered distribution. 
For instance, we can define
\begin{equation}
    \langle 0 | \O(x) \Otilde(p) | 0 \rangle
    \equiv 
    \int dy^0 dy^1 e^{-i p \cdot y} \langle 0 | \O(x) \O(y) | 0 \rangle.
\end{equation}
Adopting the convention $p^\pm = \frac{1}{2} (p^0 \pm p^1)$ such that $p \cdot y = p^+ y^+ + p^- y^-$, this is
\begin{equation}
    \langle 0 | \O(x) \Otilde(p) | 0 \rangle
    = \frac{1}{2} e^{-i p \cdot x} \Pitilde_h(p^+) \Pitilde_h(p^-).
    \label{eq:2pt:mixed}
\end{equation}
$\Pitilde_\alpha(p)$ is the Fourier transform of $\Pi_\alpha(x)$. It is equal to
\begin{equation}
    \Pitilde_\alpha(p) \equiv \int dx \, e^{i p x} \Pi_\alpha(x)
    =  \frac{2 \pi}{\Gamma(2\alpha)} R(p)^{2\alpha - 1},
\end{equation}
where $R$ is the ramp function, which obeys (with a slight abuse of notation),
\begin{equation}
    R(p)^\alpha = \left\{ \begin{array}{lll}
        p^\alpha & \text{if} & p > 0, \\
        0 & \text{if} & p \leq 0.
    \end{array}\right.
\end{equation}
The 2-point correlator vanishes therefore unless $p^\pm > 0$, or $p^0 > |p^1|$: this is the spectral condition that guarantees that only states of positive energy exist.

Eq.~\eqref{eq:2pt:mixed} can be used to compute the Fourier transform of the Gaussian correlation function \eqref{eq:4pt:Gaussian}  in $x_2$ and $x_3$.
Writing the result in terms of the sum and difference of the momenta $p_2$ and $p_3$, we obtain
\begin{equation}
    \begin{split}
        & \langle 0 | \O_a(x_1) \Otilde_a(p - q) | 0 \rangle
        \langle 0 | \Otilde_b(p + q) \O_b(x_4) | 0 \rangle 
        \\
        &= \frac{1}{4} e^{-i p \cdot (x_1 + x_4)} e^{-i q \cdot (x_4 - x_1)}
        \Pitilde_{h_a}(p^+ - q^+) \Pitilde_{\hbar_a}(p^- - q^-)
        \Pitilde_{h_b}(-p^+ - q^+) \Pitilde_{\hbar_b}(-p^- - q^-).
    \end{split}
    \label{eq:Gaussian_mixed_correlator}
\end{equation}
The next step is to take the integral of this correlator over $q^0$ as in eq.~\eqref{eq:crossing:disc}.
By the spectral condition, the integrand vanishes whenever $q^0 > -2|p^+| - q^1$ or $q^0 > -2|p^-| + q^1$. 
Inside the region where it has support, the integrand is finite.
It diverges like a power of $q^0$ in the limit $q^0 \to -\infty$, but this divergence can be regularized by adding a small imaginary part to $x_4^0$, which gives an exponential suppression. This is in fact equivalent to what happens when integrating against test functions in $x_1$ and $x_4$: the integral is finite at $|q^0| \to \infty$.
The only remaining subtlety is with the boundary point $q^0 = -2|p^+| - q^1$ or $q^0 = -2|p^-| + q^1$, whichever is smaller:
$\Pitilde_\alpha(p)$ is actually singular at $p = 0$ for $\alpha < \frac{1}{2}$, but the singularity is integrable. For generic values of $p^+$, $p^-$ and $q^1$, the integral is therefore finite and the function that it defines is analytic in $q^1$, as expected from the causality condition. However, analyticity is lost when two integrable singularities of the integrand collide: this happens at a particular value of $q^1$, defined by
\begin{equation}
    q^1_* = |p^-| - |p^+|.
\end{equation}
In this case, the integrand has support over $q^0 \in (-\infty, -|p^+| - |p^-|)$, but the limit as $q^0$ approaches the upper bound of this interval is ill-defined. As a result, the integral of eq.~\eqref{eq:Gaussian_mixed_correlator} is discontinuous around $q^1 = q^1_*$.

In order to characterize the discontinuity, we choose to work specifically with space-like $p$, and without loss of generality declare that $p^+ < 0$ and $p^- > 0$, so that $q^1_* = p^0$. We compute the integral at $q^1 = q^1_* + 2\delta$, with $|\delta| \ll |q^1_*|$, and compare the (asymptotic) limits $\delta \to 0_+$ and $\delta \to 0_-$. If the integral were analytic, the two limits would coincide. Since they turn out not to be equal, we define the discontinuity as their difference, and write
\begin{equation}
    F(x_1, x_4; p; \delta) \equiv \disc_{\delta = 0} \int dq^0 \frac{1}{q^0 - i}
    \langle 0 | \O_a(x_1) \Otilde_a(p - q) \Otilde_b(p + q) \O_b(x_4) | 0 \rangle
    \bigg|_{q^1 = q^1_* + 2\delta}.
\end{equation}
As already explained, the discontinuity arises from the boundary of the domain of integration in $q^0$, around $q^0 = p^1$. We make therefore the change of variable $q^0 = p^1 + 2t$ and focus on the regime $|t| \ll 1$. We have then
\begin{equation}
    \begin{split}
        F(x_1, x_4; p; \delta)
        & = \frac{1}{4} e^{-2i (p^- x_1^- + p^+ x_4^+)}
        \frac{\Pitilde_{\hbar_a}(2p^-) \Pitilde_{h_b}(-2p^+)}{p^1 - i}
        \\
        & \quad \times \disc_{\delta = 0} \int dt \,
        \Pitilde_{h_a}(- t- \delta) \Pitilde_{\hbar_b}(-t + \delta).
    \end{split}
\end{equation}
To compute the integral, it is convenient to express $\Pitilde_\alpha(p)$ as a sum of functions that are respectively analytic in the upper and lower half-planes,
\begin{equation}
    \Pitilde_\alpha(p) = i \Gamma(1 - 2\alpha)
    \lim_{\varepsilon \to 0_+}
    \left[ (-p + i \varepsilon)^{2\alpha - 1}
    - (-p - i \varepsilon)^{2\alpha - 1} \right].
\end{equation}
At this point, we can use the results of Appendix~\ref{sec:discontinuity} for the discontinuity of the integral, giving
\begin{equation}
    \begin{split}
        F(x_1, x_4; p; \delta)
        &= \frac{4 \pi^3 \sin \left( \pi (\hbar_b - h_a) \right)}
        {\sin\left( \pi (h_a + \hbar_b) \right)
        \Gamma(2h_a + 2\hbar_b) \Gamma(2\hbar_a) \Gamma(2h_b)}
        \\
        &\quad \times
        e^{-2i (p^- x_1^- + p^+ x_4^+)}
        \frac{R(2p^-)^{2\hbar_a - 1} R(-2p^+)^{2h_b - 1} (2\delta)^{2h_a + 2\hbar_b - 1}}{p^1 - i}.
    \end{split}
    \label{eq:F}
\end{equation}
This is a genuine discontinuity in the mixed representation of the correlation function, one that depends on the position of the operators $\O_1$ and $\O_4$ and on the combined momentum of $\O_2$ and $\O_3$.
The discontinuity of the momentum-space correlation function is obtained taking the Fourier transform with respect to $x_1$ and $x_4$. In doing so, we recover a $\delta$ function imposing overall momentum conservation. We choose therefore to define
\begin{equation}
    \int d^2x_1 d^2x_4 e^{-i(p_1 \cdot x_1 + p_4 \cdot x_4)}
    F(x_1, x_4; p; \delta)
    = (2\pi)^2 \delta^2(p_1 + 2 p + p_4)
    \widetilde{F}(p_1, p_4, \delta).
\end{equation}
In other words, $\widetilde{F}$ is the Fourier transform of $F$ modulo a 2-dimensional $\delta$ function. It is a distribution in the momenta $p_1$ and $p_4$, given by
\begin{equation}
    \begin{split}
        \widetilde{F}(p_1, p_4; p, \delta)
        &= \frac{8 \pi^5 \sin \left( \pi (\hbar_b - h_a) \right)}
        {\sin\left( \pi (h_a + \hbar_b) \right)
        \Gamma(2h_a + 2\hbar_b) \Gamma(2\hbar_a) \Gamma(2h_b)}
        \\
        &\quad \times
        \delta(p_1^+) \delta(p_4^-)
        \frac{R(-p_1^-)^{2\hbar_a - 1} R(p_4^+)^{2h_b - 1} (2\delta)^{2h_a + 2\hbar_b - 1}}{p^1 - i}.
    \end{split}
    \label{eq:Ftilde}
\end{equation}
The discontinuity is localized on the forward light cones in $-p_1$ and $p_4$ by two $\delta$ functions, specifically at $p_1^+ = p_4^- = 0$. This is a consequence of our choice to focus on the discontinuity at $q^1_* = p^0$: in general there are also discontinuities on the other sides of the cones, at $p_1^\pm = 0$ and $p_4^\pm = 0$.

The fact that this discontinuity is non-zero for generic values of the conformal weights is remarkable: by eq.~\eqref{eq:crossing:disc}, it means that there should be a similar discontinuity in the 4-point correlation function with operators $\Otilde_2$ and $\Otilde_3$ swapped.
To examine this claim, however, we need to perform more calculations: the correlation function
\begin{equation}
    \langle 0 | \O_a(x_1) \O_b(x_3) \O_a(x_2) \O_b(x_4) | 0 \rangle
\end{equation}
does not factorize into a product of 2-point functions. Instead, it admits a non-trivial expansion in conformal blocks, with intermediate operators given by composites of the form $[\O_a \partial_+^n \partial_-^{\bar{n}} \O_b]$. 
This is known as generalized free field theory.
In the next section, we show that a discontinuity at the same value of $q^1$ is found in each individual conformal block, and eventually verify the validity of the crossing equation~\eqref{eq:crossing:disc} for the Gaussian 4-point function. 
The result of the present section will also be useful in a generic CFT, whenever the identity operator enters the OPE.

%%%%%%%%%%%%%%%%%%%%%%%%%%%%%%%%%%%%%%%%%%%%%%%%%%%

\subsection{Discontinuity in the conformal blocks}
\label{sec:blocks-discontinuity}

To define a conformal block expansion for the 4-point correlation function, no matter whether in Euclidean position space or in Minkowski momentum space, we first need to specify our conventions for the OPE coefficients. We declare that the 3-point function of local primary operators satisfies
\begin{equation}
    \begin{split}
        \langle 0 | \O_1(x_1) \O_2(x_2) \O_3(x_3) | 0 \rangle
        = C_{123} \Pi_{h_{123}}(x_{12}^+) \Pi_{h_{132}}(x_{13}^+) \Pi_{h_{231}}(x_{23}^+) &
        \\
        \times \Pi_{\hbar_{123}}(x_{12}^-) \Pi_{\hbar_{132}}(x_{13}^-)
        \Pi_{\hbar_{231}}(x_{23}^-) &
    \end{split}
\end{equation}
where $h_{ijk} = h_i + h_j - h_k$ and $x_{ij} = x_i - x_j$.
$C_{123}$ is the OPE coefficient, and the rest is a conformally-covariant function of the light-cone coordinates given in terms of $\Pi_\alpha$ defined in eq.~\eqref{eq:Pi:x}.
The standard Euclidean convention is reproduced if we perform Wick rotations in the time coordinates $x_i^0$.

Note that symmetry under the exchange of the operators $\O_1$ and $\O_2$ requires that $C_{123} = (-1)^{\ell_1 + \ell_2 + \ell_3} C_{213}$, where $\ell_i$ is the spin of operator $\O_i$. Since this rule applies to any pair of operators, we have the cyclic property $C_{123} = C_{231} = C_{312}$.
Moreover, by Hermitian conjugation we must have $C_{321} = C_{123}^*$, so the OPE coefficient is real if the total spin $\ell_1 + \ell_2 + \ell_3$ is even, and purely imaginary otherwise. The symmetry property also implies that there are only even-spin operators in the OPE of two identical operators.

The Fourier transform of this 3-point function has been computed in Ref.~\cite{Gillioz:2019iye}. As before, translation symmetry implies that momentum conservation is imposed through a $\delta$ function, and we can introduce the notation
\begin{equation}
    \langle 0 | \Otilde_1(p_1) \Otilde_2(p_2) \Otilde_3(p_3) | 0 \rangle
    = (2\pi)^2 \delta^2(p_1 + p_2 + p_3)
    \llangle \Otilde_1(p_1) \Otilde_2(p_2) \Otilde_3(-p_1 - p_2) \rrangle.
\end{equation}
The object defined inside the double bracket can be factorized into left- and right-moving parts, and we have
\begin{equation}
    \begin{split}
        \llangle \Otilde_1(p_1) \Otilde_2(p_2) \Otilde_3(-p_1 - p_2) \rrangle
        \hspace{8cm} &
        \\
        = 4\pi^4 C_{123}
        R(-2p_1^+)^{h_1 + h_2 - h_3 -1}
        R(-2p_1^+ - 2p_2^+)^{2h_3-1}
        V_{h_1 h_2 h_3}\left( \tfrac{p_1^+}{p_1^+ + p_2^+} \right) &
        \\
        \times R(-2p_1^-)^{\hbar_1 + \hbar_2 - \hbar_3 -1}
        R(-2p_1^- - 2p_2^-)^{2\hbar_3-1}
        V_{\hbar_1 \hbar_2 \hbar_3}\left( \tfrac{p_1^-}{p_1^- + p_2^-} \right), &
    \end{split}
\end{equation}
where $V$ is a function defined by
\begin{equation}
    \begin{split}
        V_{h_1h_2h_3}(w)
        &= \frac{\Theta(w-1)}{\Gamma(h_1 + h_2 - h_3) \Gamma(2 h_3)}
        {}_2F_1( 1 - h_1 - h_2 + h_3; h_1 - h_2 + h_3; 2 h_3; w^{-1}) \\
        & \quad
        + \frac{\Theta(1-w) w^{h_1 - h_2 + h_3}}
        {\Gamma(h_3 + h_2 - h_1) \Gamma(2 h_1)}
        {}_2F_1( h_1 - h_2 - h_3 + 1, h_1 - h_2 + h_3; 2 h_1; w).
    \end{split}
    \label{eq:V}
\end{equation}
Since the 3-point function has support when $p_1^\pm < 0$ and $p_1^\pm + p_2^\pm < 0$, $V$ needs only to be defined for $w > 0$. The step function $\Theta$ selects which piece is relevant in the case $w \in (0, 1)$ and $w \in (1, \infty)$. In either case, the argument of the hypergeometric function is then in the unit interval $(0,1)$.

The magic of the momentum-space representation is that conformal blocks can be directly constructed as the product of 3-point functions of primary operators: among all the descendants that must be combined into a block in position space, there is a unique linear combination that carries the right momentum, and this is the one selected by the 3-point function. If we denote once again
\begin{equation}
    \langle 0 | \Otilde_1(p_1) \Otilde_2(p_2) \Otilde_3(p_3) \Otilde_4(p_4) | 0 \rangle
    = (2\pi)^2 \delta^2(p_1 + p_2 + p_3 + p_4)
    \llangle \Otilde_1(p_1) \Otilde_2(p_2) \Otilde_3(p_3) \Otilde_4(p_4) \rrangle,
\end{equation}
so that the condition $p_1 + p_2 + p_3 + p_4 = 0$ is now implicit,
then the OPE takes the form
\begin{equation}
    \begin{split}
        & \llangle \Otilde_1(p_1) \Otilde_2(p_2) \Otilde_3(p_3) \Otilde_4(p_4) \rrangle
        \\
        & \qquad\qquad
        = \sum_{\O}
        \frac{\llangle \Otilde_1(p_1) \Otilde_2(p_2) \Otilde(-p_1 - p_2) \rrangle
        \llangle \Otilde(-p_3 - p_4) \Otilde_3(p_3) \Otilde_4(p_4) \rrangle}
        {\frac{1}{2} \Pitilde_h(p_3^+ + p_4^+) \Pitilde_\hbar(p_3^- + p_4^-)}.
    \end{split}
\end{equation}
The denominator is the 2-point function of the intermediate operator $\Otilde$ that has conformal weights $(h, \hbar)$.
This expansion is valid when $p_3 + p_4$ is pointing forward, that is when $p_3^0 + p_4^0 > |p_3^1 + p_4^1|$. In all other cases the correlator vanishes by the spectral condition.
Each term in the sum is the product of a pair of OPE coefficients and of a conformal block that factorizes into left- and right-moving parts,
\begin{equation}
    \llangle \Otilde_1(p_1) \Otilde_2(p_2) \Otilde_3(p_3) \Otilde_4(p_4) \rrangle
    = \sum_\O C_{12\O} C_{43\O}^*
    G_h(p_i^+) G_\hbar(p_i^-).
\end{equation}
With the 3-point function given above, the holomorphic conformal block is 
\begin{equation}
    \begin{split}
        G_h(p_i) &= 2 \sqrt{2} \pi^3 \Gamma(2h)
        R(-p_1)^{h_1 + h_2 - h -1}
        R(p_4)^{h_4 + h_3 - h -1}
        R(p_3 + p_4)^{2h-1}
        \\
        & \quad \times
        V_{h_1 h_2 h}\left( \tfrac{p_1}{p_1 + p_2} \right)
        V_{h_4 h_3 h}\left( \tfrac{p_4}{p_3 + p_4} \right).
    \end{split}
    \label{eq:G}
\end{equation}

\subsubsection*{Convolution integral}

Now, as before, we write $p_2 = p - q$ and $p_3 = p + q$, and study the convolution integral in $q^0$ for each individual conformal block,
\begin{equation}
    \int dq^0 \frac{1}{q^0 - i} G_h(p_i^+) G_\hbar(p_i^-).
\end{equation}
We work at fixed $p_1$ and $p_4$. For the correlator to be non-zero at all, we must have $p_4$ forward-directed and $p_1$ backward-directed.
Note that momentum conservation implies $p = -\frac{1}{2}(p_1 + p_4)$, so the only remaining variable in this integral is $q^1$.

The first observation that we can make is that the conformal block has support when $p_3 + p_4$ is forward-directed.
Since $p_3 + p_4 = q + \frac{1}{2}(p_4 - p_1)$, this gives a lower bound on $q^0$, below which the integrand vanishes.
Above this bound, the behavior of the integrand is dictated by the function $V$ defined in eq.~\eqref{eq:V}. There are 3 special cases corresponding to the limits $w \to 0_+$, $w \to 1$, and $w \to \infty$. In all other regimes the integrand is smooth.

\subsubsection*{Small exchange momentum}

The case $w \to \infty$ is attained when the exchange momentum $p_3 + p_4$ (and hence $p_1 + p_2$) is light-like, or equivalently when the center-of-mass energy of the $2 \to 2$ scattering process vanishes. This corresponds to the lower bound of the convolution integral in $q^0$.
For generic values of $q^1$, either $p_3^+ + p_4^+ \to 0_+$ or $p_3^- + p_4^- \to 0_+$, but not both simultaneously. Let us examine the former case:
since $V(w)$ is finite in the limit $w \to \infty$, the behavior of the integrand is determined by $R(p_3^+ + p_4^+)^{2h -1}$, and the convolution integral of the conformal block is therefore convergent for all $h > 0$.

If $q^1$ is such that both $p_3^\pm + p_4^\pm \to 0$ at the lower boundary in $q^0$, then the integral diverges for operators that have $\Delta = h + \hbar \leq 1$. It can be defined by analytic continuation from larger values of $\Delta$, but we may as well choose to stay away from this point. 

\subsubsection*{Large exchange momentum}

The limit $q^0 \to \infty$ of the convolution integral corresponds to a large center-of-mass energy, and it is given by the asymptotic limit $w \to 0_+$ of the function $V(w)$, satisfying 
\begin{equation}
    V_{h_1h_2h}(w) \propto |w|^{h_1 - h_2 + h}
    \qquad\qquad
    \text{as} ~ w \to 0_+.
\end{equation}
When combining it with the other elements of the holomorphic block, we get
\begin{equation}
    G_h(p_i) \propto
    (q^0)^{h_2 + h_3 - h_1 - h_4 - 1}
    \qquad\qquad
    \text{as} ~ q^0 \to \infty.
\end{equation}
The whole conformal block satisfies therefore
\begin{equation}
    G_h(p_i^+) G_\hbar(p_i^-)
    \propto (q^0)^{\Delta_2 + \Delta_3 - \Delta_1 - \Delta_4 - 2}
    \qquad\qquad
    \text{as} ~ q^0 \to \infty,
\end{equation}
and the convolution integral is convergent if
\begin{equation}
    \Delta_2 + \Delta_3 < 2 + \Delta_1 + \Delta_4.
    \label{eq:convergencecondition}
\end{equation}
We restrict our analysis to correlation functions of operators satisfying this condition. This includes the case of identical operators.

\subsubsection*{Light cone crossing}

\begin{figure}
    \centering
    \begin{subfigure}{0.32\linewidth}
        \includegraphics[width=\linewidth]{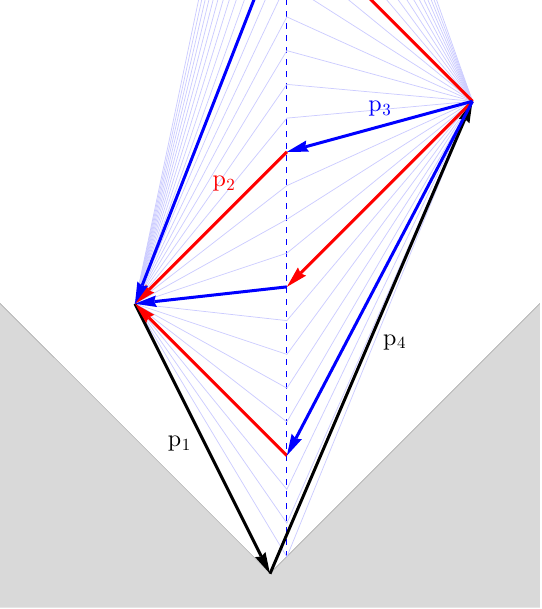} 
        \caption{Regular case}
        \label{fig:q-integral-regular}
    \end{subfigure}
    \begin{subfigure}{0.32\linewidth}
        \includegraphics[width=\linewidth]{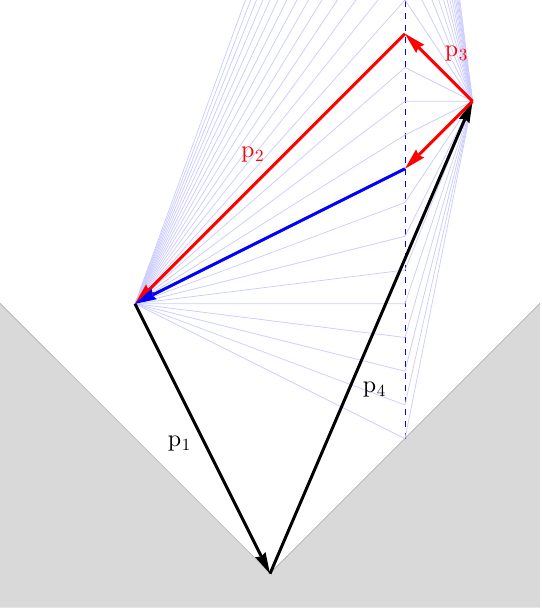} 
        \caption{s-channel point}
        \label{fig:q-integral-s}
    \end{subfigure}
    \begin{subfigure}{0.32\linewidth}
        \includegraphics[width=\linewidth]{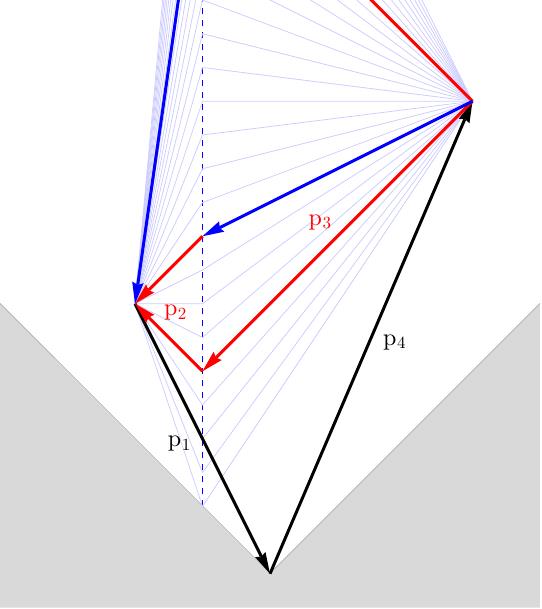} 
        \caption{t-channel point}
        \label{fig:q-integral-t}
    \end{subfigure}
    \caption{Configurations of four momenta $p_1$ to $p_4$ adding up to zero, $p_1 + p_2 + p_3 + p_4 = 0$, with different values of $q = p_2 - p_3$. In each case (a), (b), and (c), the spatial component $q^1$ (along the horizontal direction) is held fixed and the time component $q^0$ varies (along the vertical direction). In the generic case (a), $p_2$ or $p_3$ can be light-like, as indicated with a red arrow, but never simultaneously.
    There are two particular values of $q^1$ at which $p_2$ and $p_3$ are both light-like, shown in (b) and (c). These two cases are related to each other by the exchange $p_2 \leftrightarrow p_3$, or $q \leftrightarrow -q$.
    Configurations of the type (a), (b) and (c) are found in any situation in which the momenta $-p_1$ and $p_4$ are forward-directed (outside the gray region) but neither colinear nor separated by a light-like vector.}
    \label{fig:q-integral}
\end{figure}

The limit $w \to 1$ is reached at several points in the bulk of the integration domain, whenever $p_2$ or $p_3$ becomes light-like, as illustrated in Figure~\ref{fig:q-integral}.
Eq.~\eqref{eq:V} is not particularly well-suited to understand this limit. Instead, we can use hypergeometric identities to give a different representation of the same function:
\begin{equation}
    \begin{split}
        V_{h_1h_2h}(w)
        &= \frac{\Gamma(2h_2 - 1)}{\Gamma(h_1 + h_2 + h - 1)
        \Gamma(h_1 + h_2 - h) \Gamma(-h_1 + h_2 + h)} \\
        & \quad\qquad\times
        {}_2F_1( 1 - h_1 - h_2 + h, h_1 - h_2 + h; 2 - 2 h_2; 1 - w^{-1}) \\
        & \quad
        + \frac{\Gamma(1 - 2 h_2)}{\Gamma(h_1 - h_2 + h)}
        \left[ \frac{e^{-i \pi (h_1 + h_2 - h)}}{2 \pi}
        (1 - w^{-1} + i \varepsilon)^{2h_2 -1} + \text{c.c.} \right] \\
        & \quad\qquad\times
        {}_2F_1( h_1 + h_2 + h - 1, -h_1 + h_2 + h; 2 h_2; 1 - w^{-1}),
    \end{split}
    \label{eq:V:alt}
\end{equation}
where the limit $\varepsilon \to 0_+$ is understood as usual. The first piece of this function (first two lines) is analytic in $w$ around $w = 1$. The second piece is not: in fact, it is even singular for $h_2 < \frac{1}{2}$, but the singularity is integrable.
As indicated, $V$ is the sum of two functions that are each analytic in the upper and lower half-plane respectively. In the generic case illustrated in Fig.~\ref{fig:q-integral-regular}, this means that the integration contour in $q^0$ can be deformed inside the domain where the integrand is smooth.
The convolution integral is therefore finite and defines an analytic function of $q^1$.

\subsubsection*{Double light cone crossing}

There are however special values of $q^1$ such that both 3-point functions in the conformal block are simultaneously singular at some value of $q^0$, as illustrated in Figs.~\ref{fig:q-integral-s} and \ref{fig:q-integral-t}. When this happens, the previous argument for analyticity of the convolution integral fails: the integration contour cannot be deformed into the complex plane because it is pinched between singularities on both sides.

Like we did with the Gaussian correlator above, we choose to work with a momentum $p = -\frac{1}{2}(p_1 + p_4)$ that is space-like, with $p^+ < 0$ and $p^- > 0$, such that
\begin{equation}
    0 < -p_1^+ < p_4^+,
    \qquad\text{and}\qquad
    0 < p_4^- < -p_1^-.
\end{equation}
With this choice, the two special values are $q^1 = \pm q^1_*$, defined by 
\begin{equation}
    q^1_* = p^0 = -\frac{p_1^+ + p_4^+ + p_1^- + p_4^-}{2}.
\end{equation}
At these values, the integral crosses two light cone simultaneously at $q^0 = \pm q^0_*$, with
\begin{equation}
    q^0_* = p^1 = -\frac{p_1^+ + p_4^+ - p_1^- - p_4^-}{2}.
\end{equation}
The two special values are related by $q \leftrightarrow -q$, or the exchange $p_2 \leftrightarrow p_3$.
We call $q^1 = - q_*^1$ the s-channel point, and $q^1 = + q_*^1$ the t-channel point.

\subsubsection*{s-channel discontinuity}
At the s-channel point, we have
\begin{equation}
    p_2^- = p_3^+ = 0,
    \qquad
    p_2^+ = -p_1^+ - p_4^+ < 0,
    \qquad
    p_3^- = -p_1^- - p_4^- > 0
    \qquad
    \text{(s channel).}
\end{equation}
In a neighborhood of this point, we can write $q^1 = -q^1_* + 2\delta$, $q^0 = -q^0_* + 2t$, and work under the assumption that $|\delta|, |t| \ll 1$. We have then $p_2^- \approx \delta - t$ and $p_3^+ \approx \delta + t$, and the conformal blocks obey
\begin{align}
    G_h(p_i^+) &\approx 2^{3/2} \pi^3 \Gamma(2h)
    (-p_1^+)^{h_1 + h_2 - h - 1}
    (p_4^+)^{h_4 + h_3 + h - 2}
    V_{h_1 h_2 h}\left( -\tfrac{p_1^+}{p_4^+} \right)
    V_{h_4 h_3 h}\left( \tfrac{p_4^+}{p_4^+ + \delta + t} \right)
    \\
    G_\hbar(p_i^-) &\approx 2^{3/2} \pi^3 \Gamma(2\hbar)
    (-p_1^-)^{\hbar_1 + \hbar_2 + \hbar - 2}
    (p_4^-)^{\hbar_4 + \hbar_3 - \hbar - 1}
    V_{\hbar_1 \hbar_2 \hbar}\left( \tfrac{p_1^-}{p_1^- + \delta - t} \right)
    V_{\hbar_4 \hbar_3 \hbar}\left( - \tfrac{p_4^-}{p_1^-} \right).
\end{align}
Two of the functions $V$ take values at regular arguments, which we denote
\begin{equation}
    w = -\frac{p_1^+}{p_4^+},
    \qquad\qquad
    \wbar = -\frac{p_4^-}{p_1^-}
    \qquad\qquad
    w, \wbar \in (0, 1).
    \label{eq:wwbar}
\end{equation}
The other two must be evaluated at argument one using eq.~\eqref{eq:V:alt}.
Neglecting pieces that are analytic in $\delta$, this gives 
\begin{align}
    G_h(p_i^+) &\approx 
    \frac{\sqrt{2} \pi^2 \Gamma(2h) \Gamma(1 - 2 h_3)}{\Gamma(h_4 - h_3 + h)}
    (p_4^+)^{h_1 + h_2 - h_3 + h_4 - 2}
    w^{h_1 + h_2 - h - 1}
    V_{h_1 h_2 h}(w)
    \nonumber \\
    & \quad \times
    \left[ e^{-i \pi (h_3 + h_4 - h)}
    (-t - \delta + i \varepsilon)^{2h_3 -1} + \text{c.c.} \right],
    \\
    G_\hbar(p_i^-) &\approx 
    \frac{\sqrt{2} \pi^2 \Gamma(2\hbar) \Gamma(1 - 2 \hbar_2)}
    {\Gamma(\hbar_1 - \hbar_2 + \hbar)}
    (-p_1^-)^{\hbar_1 - \hbar_2 + \hbar_3 + \hbar_4 - 2}
    \wbar^{\hbar_4 + \hbar_3 - \hbar - 1}
    V_{\hbar_4 \hbar_3 \hbar}(\wbar)
    \nonumber \\
    & \quad \times
    \left[ e^{-i \pi (\hbar_1 + \hbar_2 - \hbar)}
    (- t + \delta + i \varepsilon)^{2\hbar_2 - 1} + \text{c.c.} \right].
\end{align}
Multiplying these two functions and integrating over $t$, one recovers a situation observed with the Gaussian 4-point function: there is a discontinuity at $\delta = 0$.
The results of Appendix \ref{sec:discontinuity} tell us that
\begin{equation}
    \begin{split}
        \disc_{q^1 = -q^1_*} & \int dt
        \Big[ e^{-i \pi (h_3 + h_4 - h)}
        (-t - \delta + i \varepsilon)^{2h_3 -1} + \text{c.c.} \Big]
        \Big[ e^{-i \pi (\hbar_1 + \hbar_2 - \hbar)}
        (- t + \delta + i \varepsilon)^{2\hbar_2 - 1} + \text{c.c.} \Big]
        \\
        &= \frac{4 \pi \sin \left[ \pi (\hbar_1 - h_4 + h - \hbar) \right]
        (2\delta)^{2h_3 + 2\hbar_2 - 1}}
        {\sin\left[ \pi (h_3 + \hbar_2) \right] \Gamma(2h_3 + 2\hbar_2)
        \Gamma(1 - 2h_3) \Gamma(1 - 2\hbar_2)}.
    \end{split}
\end{equation}
Putting the pieces together, we obtain an expression for the discontinuity of the convolution integral at the s-channel point,
\begin{equation}
    \begin{split}
        & \disc_{q^1 = -q^1_*} \int dq^0 \frac{1}{q^0 - i} G_h(p_i^+) G_\hbar(p_i^-)
        \\
        & \quad 
        = \frac{(-1)^{\ell + \ell_1} \Gamma(2h) \Gamma(2\hbar)}
        {\Gamma(h_4 - h_3 + h) \Gamma(\hbar_1 - \hbar_2 + \hbar)}
        \frac{8 \pi^5 \sin\left[ \pi (h_1 - h_4) \right]}
        {\sin\left[ \pi (h_3 + \hbar_2) \right] \Gamma(2h_3 + 2\hbar_2)}
        \frac{(2\delta)^{2h_3 + 2\hbar_2 - 1}}{-q^0_* - i}
        \\
        & \quad\quad \times
        (p_4^+)^{h_1 + h_2 - h_3 + h_4 - 2}
        (-p_1^-)^{\hbar_1 - \hbar_2 + \hbar_3 + \hbar_4 - 2}
        w^{h_1 + h_2 - h - 1}
        V_{h_1 h_2 h}(w)
        \wbar^{\hbar_4 + \hbar_3 - \hbar - 1}
        V_{\hbar_4 \hbar_3 \hbar}(\wbar).
    \end{split}
    \label{eq:discontinuity:s}
\end{equation}
Note that we have used the fact that $h_i - \hbar_i = \ell_i$ is integer to write
\begin{equation}
    \sin \left[ \pi (\hbar_1 - h_4 + h - \hbar) \right]
    = (-1)^{\ell + \ell_1} \sin \left[ \pi (h_1 - h_4) \right].
\end{equation}

\subsubsection*{t-channel discontinuity}
At the t-channel point, we have
\begin{equation}
    p_2^+ = p_3^- = 0,
    \qquad
    p_2^- = -p_1^- - p_4^- > 0,
    \qquad
    p_3^+ = -p_1^+ - p_4^+ < 0
    \qquad
    \text{(t channel)}.
\end{equation}
As before, we write $q^1 = q^1_* + 2\delta$ and $q^0 = q^0_* + 2t$, and examine the conformal blocks in the neighborhood $|\delta|, |t| \ll 1$, giving
\begin{align}
    G_h(p_i^+) &\approx 2^{3/2} \pi^3 \Gamma(2h)
    (-p_1^+)^{h_1 + h_2 + h - 2}
    (p_4^+)^{h_4 + h_3 - h - 1}
    V_{h_1 h_2 h}\left( \tfrac{p_1^+}{p_1^+ - \delta - t} \right)
    V_{h_4 h_3 h}\left( -\tfrac{p_4^+}{p_1^+} \right),
    \\
    G_\hbar(p_i^-) &\approx 2^{3/2} \pi^3 \Gamma(2\hbar)
    (-p_1^-)^{\hbar_1 + \hbar_2 - \hbar - 1}
    (p_4^-)^{\hbar_4 + \hbar_3 + \hbar - 2}
    V_{\hbar_1 \hbar_2 \hbar}\left( -\tfrac{p_1^-}{p_4^-} \right)
    V_{\hbar_4 \hbar_3 \hbar}\left( \tfrac{p_4^-}{p_4^- - \delta + t} \right).
\end{align}
Proceeding as before, we obtain
\begin{equation}
    \begin{split}
        & \disc_{q^1 = q^1_*} \int dq^0 \frac{1}{q^0 - i} G_h(p_i^+) G_\hbar(p_i^-)
        \\
        &= \frac{(-1)^{\ell + \ell_4} \Gamma(2h) \Gamma(2\hbar)}
        {\Gamma(h_1 - h_2 + h) \Gamma(\hbar_4 - \hbar_3 + \hbar)}
        \frac{8 \pi^5 \sin\left[ \pi (h_1 - h_4) \right]}
        {\sin\left[ \pi (h_2 + \hbar_3) \right]
        \Gamma(2h_2 + 2\hbar_3)}
        \frac{(2\delta)^{2h_2 + 2\hbar_3 - 1}}{q^0_* - i}
        \\
        & \quad \times
        (p_4^+)^{h_1 - h_2 + h_3 + h_4 - 2}
        (-p_1^-)^{\hbar_1 + \hbar_2 - \hbar_3 + \hbar_4 - 2}
        w^{h_1 - h_2 + h - 1}
        V_{h_4 h_3 h}(w^{-1})
        \wbar^{\hbar_4 - \hbar_3 + \hbar - 1}
        V_{\hbar_1 \hbar_2 \hbar}(\wbar^{-1}).
    \end{split}
    \label{eq:discontinuity:t}
\end{equation}
Note that the arguments of the functions $V$ are inverted in comparison with the s-channel point.

%%%%%%%%%%%%%%%%%%%%%%%%%%%%%%%%%%%%%%%%%%%%%%%%%%%

\subsection{A new crossing equation in 2d}

Since we found non-zero discontinuities in each conformal block, we must be able to write a crossing equation in the form of eq.~\eqref{eq:crossing:disc}, relating distinct OPEs of the 4-point function. At this stage, we cannot exclude the possibility that the computation of the discontinuity does not commute with the infinite sum over intermediate operators. However, we will see below that this is not the case.

Let us focus on the discontinuity at the s-channel point. From the result of eq.~\eqref{eq:discontinuity:s}, we can write
\begin{equation}
    \disc_{q^1 = -q^1_*} \int dq^0 \frac{1}{q^0 - i}
    \llangle \Otilde_1(p_1) \Otilde_2(p - q) \Otilde_3(p + q) \O_4(p_4) \rrangle
    = Z \sum_{\O} C_{12\O} C_{34\O}^* S_h(w) S_\hbar(\wbar).
\end{equation}
where we have used the property $C_{43\O} = (-1)^{\ell_3 + \ell_4 + \ell} C_{34\O}$ and the fact that the total spin $\ell_1 + \ell_2 + \ell_3 + \ell_4$ is an even integer to hide the factor of $(-1)^\ell$.
Here $Z$ is a quantity common to all conformal blocks, independent of the conformal weights $h$, $\hbar$ of the internal operator, which we define as
\begin{equation}
    Z = \frac{8 \pi^5 \sin \left[ \pi (h_1 - h_4) \right] (-1)^{\ell_2}}
    {\sin\left( \pi (h_3 + \hbar_2) \right)
    \Gamma(2h_3 + 2\hbar_2)}
    \frac{(2\delta)^{2h_3 + 2\hbar_2 - 1}}{q^0_* + i}
    (p_4^+)^{h_1 + h_2 - h_3 + h_4 - 2}
    (-p_1^-)^{\hbar_1 - \hbar_2 + \hbar_3 + \hbar_4 - 2}.
    \label{eq:Z}
\end{equation}
With this choice, the rest only depends on the ratio of momenta $w$ and $\wbar$ defined in eq.~\eqref{eq:wwbar}, and factorizes into holomorphic and anti-holomorphic parts, 
\begin{align}
    S_h(w) &= \frac{\Gamma(2h)}{\Gamma(h_4 - h_3 + h)}
    w^{h_1 + h_2 - h - 1}
    V_{h_1 h_2 h}(w),
    \\
    S_\hbar(\wbar) &= \frac{\Gamma(2\hbar)}{\Gamma(\hbar_1 - \hbar_2 + \hbar)}
    \wbar^{\hbar_4 + \hbar_3 - \hbar - 1}
    V_{\hbar_4 \hbar_3 \hbar}(\wbar).
\end{align}
This discontinuity at the s-channel point must be equal to the discontinuity of the crossed correlator at the t-channel point, since the two are related by the joint exchange $\Otilde_2 \leftrightarrow \Otilde_3$ and $q \leftrightarrow -q$.
Repeating the same procedure starting from eq.~\eqref{eq:discontinuity:t}, we get
\begin{equation}
    \disc_{q^1 = -q^1_*} \int dq^0 \frac{1}{q^0 - i}
    \llangle \Otilde_1(p_1) \Otilde_3(p + q) \Otilde_2(p - q) \O_4(p_4) \rrangle
    = Z \sum_{\O} C_{13\O} C_{24\O}^* T_h(w) T_\hbar(\wbar),
\end{equation}
where
\begin{align}
    T_h(w) &= \frac{\Gamma(2h)}{\Gamma(h_1 - h_3 + h)}
    w^{h_1 - h_3 + h - 1}
    V_{h_4 h_2 h}(w^{-1}),
    \\
    T_\hbar(\wbar) &= \frac{\Gamma(2\hbar)}{\Gamma(\hbar_4 - \hbar_2 + \hbar)}
    \wbar^{\hbar_4 - \hbar_2 + \hbar - 1}
    V_{\hbar_1 \hbar_3 \hbar}(\wbar^{-1}).
\end{align}
Equating the two discontinuities gives finally the crossing equation
\begin{equation}
    \sum_{\O} C_{12\O} C_{34\O}^* S_h(w) S_\hbar(\wbar)
    = \sum_{\O} C_{13\O} C_{24\O}^* T_h(w) T_\hbar(\wbar).
    \label{eq:2dcrossing}
\end{equation}
Writing the holomorphic blocks in explicit form, we find
\begin{align}
    S_h(w) &= \frac{\Gamma(2h) w^{2h_1 - 1}}
    {\Gamma(h_4 - h_3 + h) \Gamma(h_2 - h_1 + h) \Gamma(2 h_1)}
    {}_2F_1( h_1 - h_2 - h + 1, h_1 - h_2 + h; 2 h_1; w),
    \label{eq:S}
    \\
    T_h(w) &= \frac{w^{h_1 - h_3 + h - 1}}
    {\Gamma(h_1 - h_3 + h) \Gamma(h_2 + h_4 - h)}
    {}_2F_1( 1 - h_2 - h_4 + h; h_4 - h_2 + h; 2 h; w).
    \label{eq:T}
\end{align}
The anti-holomorphic blocks are obtained replacing $h \to \hbar$ and $h_i \to \hbar_{5-i}$. For convenience, we will refer to $S_h$ as the \emph{s-channel conformal block}, and to $T_h$ as the \emph{t-channel block}.
This distinction is necessary because they are genuinely distinct functions, unlike the ordinary position-space conformal block that have built-in crossing symmetry.

\subsubsection*{Pairwise-identical operators}

In most practical cases, we do not consider 4 distinct external operators $\O_1$, $\O_2$, $\O_3$, and $\O_4$, but rather pairs of distinct operators, or even identical operators as in the introduction. These situations are particularly interesting in a bootstrap perspective because they make squares of the OPE coefficients appear on one or the other side of the crossing equation, with subsequent positivity properties.

There are 3 possible pairings of operators.
In the first two cases, the sets of operators entering the sum in the s and t channel are in principle distinct:
\begin{equation}
    \O_1 = \O_2, ~ \O_3 = \O_4: \qquad
    \sum_{\O} C_{11\O} C_{44\O} S_h(w) S_\hbar(\wbar)
    =\sum_{\O} \left| C_{14\O} \right|^2  T_h(w) T_\hbar(\wbar),
    \label{eq:2dcrossing:identical12}
\end{equation}
\begin{equation}
    \O_1 = \O_3, ~ \O_2 = \O_4: \qquad
    \sum_{\O} \left| C_{14\O} \right|^2 S_h(w) S_\hbar(\wbar)
    = \sum_{\O} C_{11\O} C_{44\O} T_h(w) T_\hbar(\wbar).
    \label{eq:2dcrossing:identical13}
\end{equation}
Note that $C_{11\O}$ and $C_{44\O}$ are real.
In the third case, we are tempted to write an equation that involves the same sum over operators in both channels:
\begin{equation}
    \O_1 = \O_4, ~ \O_2 = \O_3: \qquad
    \sum_{\O} C_{12\O}^2 
    \left[ S_h(w) S_\hbar(\wbar) - T_h(w) T_\hbar(\wbar) \right] \stackrel{?}{=} 0.
    \label{eq:2dcrossing:identical14}
\end{equation}
However, this equation is incorrect.
We will see later in Section~\ref{sec:newbasis} that the left-hand side is generically non-zero: in generalized free field theory, there exists a functional that isolates the contribution of a single operator in the sum.
The reason for this failure is the vanishing of the discontinuity when $h_1 = h_4$: in defining the crossing equation and the conformal blocks $S_h$ and $T_h$, we have assumed the finiteness of the factor $Z$ of eq.~\eqref{eq:Z}, but it is not true when $\O_1 = \O_4$. What happens in this case is that the discontinuity vanishes in both correlators and there is no crossing equation.

This is a surprising feature. In fact, if we take it for granted, we cannot even consider the case of identical operators mentioned in the introduction and successfully put to the test below.
The resolution of this problem is to consider the case of identical operators as a limit.
This can work if the limit does not require introducing operators in the OPE that are otherwise forbidden by symmetry.
As a matter of fact, both eqs.~\eqref{eq:2dcrossing:identical12} and \eqref{eq:2dcrossing:identical13} can be used to recover the crossing equation \eqref{eq:2dcrossing:identical} in the limit $h_1, h_4, \hbar_1, \hbar_4 \to h_\phi$, but in both equations the identity operator is missing on one side when $\O_1 \neq \O_4$.

\subsubsection*{The identity operator}

We have so far considered intermediate operators with positive conformal weights $h, \hbar > 0$. But some correlation functions also include the identity operator with $h = \hbar = 0$ in their OPE: this happens when the operators are pairwise-identical, as in eqs.~\eqref{eq:2dcrossing:identical12} and \eqref{eq:2dcrossing:identical13}.

The identity block is given by discontinuity of the the Gaussian correlation function \eqref{eq:4pt:Gaussian}. Looking back at the result of Section~\ref{sec:Gaussianexample}, we see that there is no discontinuity at the s-channel point $q^1 = -q^1_*$. This is actually consistent with the vanishing of the s-channel block $S_h$ when the conformal weights $h$ and $\hbar$ are taken to zero after setting $h_1 = h_2$ and $h_3 = h_4$:
\begin{equation}
    S_0(w) = \lim_{h \to 0} S_h(w) = 0.
    \label{eq:S:0}
\end{equation}
The Gaussian correlator does however have a discontinuity at the t-channel point: its value was computed and is given in eq.~\eqref{eq:Ftilde}. Dividing it by the conventional factor $Z$ of eq.~\eqref{eq:Z} and splitting the results into holomorphic and anti-holomorphic pieces, we obtain
\begin{equation}
    T_0(w) = \frac{\delta(w)}{\Gamma(2h_4)},
    \qquad\qquad
    T_0(\wbar) = \frac{\delta(\wbar)}{\Gamma(2\hbar_1)}.
    \label{eq:T:0}
\end{equation}
This is again consistent with the limit $h \to 0_+$ of the block $T_h(w)$ after setting $h_1 = h_3$ and $h_2 = h_4$. Indeed, the limit is identically zero at any $w > 0$, and in a neighborhood of the point $w = 0$, we have
\begin{equation}
    T_h(w) \sim \frac{\Theta(w) w^{h - 1}}
    {\Gamma(h) \Gamma(2h_4)}
    \qquad \text{as} ~ w \to 0.
\end{equation}
The subtlety here is that we need to consider $T_h(w)$ to be a measure over $w \in \mathbb{R}$, and not just on the interval $w \in (0,1)$, hence the presence of the step function in the numerator. Then we can rewrite this as
\begin{equation}
    T_h(w) \sim \frac{i \Gamma(1 - h)}
    {2\pi \Gamma(2h_4)}
    \lim_{\varepsilon \to 0_+}
    \left[ e^{-i \pi h} (w + i \varepsilon)^{h-1}
    - e^{i \pi h} (w - i \varepsilon)^{h-1} \right]
    \qquad \text{as} ~ w \to 0,
\end{equation}
which has a perfectly well-defined limit $h \to 0$, giving
\begin{equation}
    T_0(w) = \frac{i}
    {2\pi \Gamma(2h_4)}
    \lim_{\varepsilon \to 0_+}
    \left[ \frac{1}{w + i \varepsilon}
    -  \frac{1}{w - i \varepsilon} \right].
\end{equation}
We recognize on the right-hand side a well-known representation of the $\delta$ function, in agreement with eq.~\eqref{eq:T:0}.

The case of (anti-)holomorphic intermediate operators can be treated similarly to the identity: when $h = 0$ (respectively $\hbar = 0$), the corresponding (anti-)holomorphic part of the conformal block is given by eqs.~\eqref{eq:S:0} and \eqref{eq:T:0}, and the other part by eq.~\eqref{eq:S} and \eqref{eq:T}.

\subsubsection*{Identical external operators}

Since the contribution of the identity operator is zero in the s channel, it must be possible to take the limit of identical operators starting from the case $\O_1 = \O_3$ and $\O_2 = \O_4$ given in eq.~\eqref{eq:2dcrossing:identical13} and taking the limit $\O_4 \to \O_1$, without omitting any operator in the OPE.
What happens is instead the opposite: there are odd-spin operators present in the case $\O_1 \neq \O_4$ that cannot appear when $\O_1 = \O_4$ by symmetry.
Since these operators do not go away in the limit $\O_1 \to \O_4$, any solution to crossing derived from the momentum-space equation will include fictitious operators that do not exists in a physical theory.

This apparent paradox is resolved by the fact that there are actually two solutions to crossing in a neighborhood of the identical limit: one of them determines the OPE coefficients $C_{14\O}$, the other $C_{41\O} = (-1)^\ell C_{14\O}$. The t-channel side of the crossing equation \eqref{eq:2dcrossing:identical13} is the same in both cases.
Since both equations are valid, we can consider their symmetric and antisymmetric combinations. Their difference implies that the sum over odd-spin internal operators in the s channel is zero, which is consistent with the vanishing of the OPE coefficients. The symmetric combination implies then
\begin{equation}
    C_{\phi\phi\O}^2 = \left[ 1 + (-1)^\ell \right] \lim_{\O_1, \O_4 \to \phi} \left|C_{14\O} \right|^2.
    \label{eq:identicallimit}
\end{equation}
This is what is observed in generalized free field theory~\cite{Fitzpatrick:2011dm, Fitzpatrick:2012yx}.
Apart from this subtlety, the conformal blocks \eqref{eq:S} and \eqref{eq:T} have an unambiguous limit when all operators are identical, corresponding to eqs.~\eqref{eq:S:identical} and \eqref{eq:T:identical} presented in the introduction.

\subsubsection*{Generalized free field theory}

With these results, we are now armed to go back to the example of Section~\ref{sec:Gaussianexample}, distinguishing 3 cases:

\begin{itemize}
    \item 
    For a Gaussian correlator with $\O_1 = \O_2$ and $\O_3 = \O_4$, as in eq.~\eqref{eq:4pt:Gaussian}, the equation is trivially satisfied: there is only the identity in the s-channel, and its contribution vanishes; in the t channel there are double trace operator of the schematic form $[\O_1 \partial_+^n \partial_-^{\bar{n}} \O_4]$, with conformal weights
    $h = h_1 + h_4 + n$ and $\hbar = \hbar_1 + \hbar_4 + \bar{n}$, but the t-channel block has zeros precisely at these values, due to a $\Gamma$ function with negative integer argument in the denominator.

    \item 
    For a Gaussian correlator with $\O_1 = \O_3$ and $\O_2 = \O_4$, the s-channel sum has an infinite tower of composite operators, while there is just the identity operator in the t channel. Both sides of the equation are non-zero, and we will be able to verify its validity in Section~\ref{sec:newbasis} below.

    \item 
    Finally, in the case $\O_1 = \O_4$ and $\O_2 = \O_3$, the discontinuity vanishes in both correlators and there is no crossing equation.
\end{itemize}

\subsection{The crossing equation in 1d}

Consider the case of a holomorphic operator $\chi$ that only depend on the position $x^+$ and is independent of $x^-$. It can be used to define a correlation function on the line,
\begin{equation}
    \langle 0 | \chi(x_1^+) \chi(x_2^+) \chi(x_3^+) \chi(x_4^+) | 0 \rangle.
\end{equation}
$\chi$ is a genuine 2-dimensional operator with zero anti-holomorphic weight $\hbar_\chi = 0$. 
The correlator admits a conformal block expansion with internal operators that are also holomorphic, with $\hbar = 0$.

At first, it seems impossible to apply the momentum-space crossing equation to this one-dimensional system: it relies very much on two dimensional kinematics and cannot be localized on a line. Indeed, if we attempt to take the limit $\hbar_i \to 0_+$ of the previous results, we do not obtain anything meaningful:
\begin{itemize}
    \item If we take $\hbar_3 \to 0$, the internal operator must have $\hbar = \hbar_4$ in the s channel and $\hbar = \hbar_1$ in the t channel. In both $S_\hbar$ and $T_\hbar$ the hypergeometric function remains finite in this limit, but the argument of one of the $\Gamma$ functions in the denominator goes to zero, so that the whole block vanishes. The crossing equation is therefore trivially satisfied.

    \item The same situation happens if we send $\hbar_1 \to 0$ first: we must have $\hbar = \hbar_2$ in the s channel and $\hbar = \hbar_3$ in the t channel, and in both cases the blocks vanish.
\end{itemize}
However, note that the other two limits are more interesting: 
\begin{itemize}
    \item When $\hbar_2 \to 0$, with $\hbar = \hbar_1$ in the s channel and $\hbar = \hbar_4$ in the t channel, the two conformal blocks are equal and satisfy
    \begin{equation}
        S_\hbar(\wbar) = T_\hbar(\wbar) = \frac{\wbar^{2\hbar_4 - 1}}
        {\Gamma(\hbar_1 + \hbar_3 - \hbar_4) \Gamma(2 \hbar_4)}
        {}_2F_1( 1 - \hbar_1 - \hbar_3 + \hbar_4, \hbar_1 - \hbar_3 + \hbar_4; 2 \hbar_4; \wbar).
    \end{equation}

    \item Similarly, in the limit $\hbar_4 \to 0_+$ with $\hbar = \hbar_3$ in the s channel and $\hbar = \hbar_2$ in the t channel, we find that the two blocks are identical and obey 
    \begin{equation}
        S_\hbar(\wbar) = T_\hbar(\wbar) = \frac{\delta(\wbar)}{\Gamma(\hbar_1 - \hbar_2 + \hbar_3)},
    \end{equation}
    where we have used the identity encountered above
    \begin{equation}
        \lim_{\alpha \to 0_+} \alpha R(\wbar)^{\alpha -1} = \delta(\wbar).
    \end{equation}
\end{itemize}
This suggests the following construction: let us add a spectator anti-holomorphic field $\bar{\chi}_1(x^-)$ to the holomorphic theory, with conformal weights $(h, \hbar) = (0, \hbar_1)$, and consider the crossing equation involving the composite operators
\begin{equation}
    \mathcal{O}_1 = \mathcal{O}_3 = [\chi_1 \bar{\chi}_1],
    \qquad
    \mathcal{O}_2 = \mathcal{O}_4 = \chi_2,
\end{equation}
where $\chi_1$ and $\chi_2$ are two holomorphic operators, possibly but not necessarily distinct.
Given that the operators are pair-wise identical, we can make use of eq.~\eqref{eq:2dcrossing:identical13}.
The two anti-holomorphic blocks are equal in the joint limit $\hbar_2, \hbar_4 \to 0_+$,
\begin{equation}
    S_\hbar(\wbar) = T_\hbar(\wbar) = \frac{\delta(\wbar)}{\Gamma(2\hbar_1)}.
\end{equation}
Since the equation is trivial in the anti-holomorphic sector, 
the following holomorphic equation must hold:
\begin{equation}
    \sum_{\O} \left| C_{12\O} \right|^2 S_h(w)
    = \sum_{\O} C_{11\O} C_{22\O} T_h(w).
    \label{eq:1dcrossing}
\end{equation}
This is the one-dimensional momentum-space crossing equation.

%%%%%%%%%%%%%%%%%%%%%%%%%%%%%%%%%%%%%%%%%%%%%%%%%%%

\subsection{Convergence}

An important question regarding the crossing equation that we have derived is the convergence of the infinite sum on either side. We expect from first principles of quantum field theory that the momentum-space conformal block expansion converges in the distributional sense. In fact, it was shown in Ref.~\cite{Gillioz:2019iye} that the convergence is even point-wise in a certain kinematic regime.
This happens in the t channel, but not in the s channel.
We quickly review the argument here and provide an actual example.

The asymptotics of the s- and t-channel blocks at large $h$ are quite different.
On the one hand, we have 
\begin{equation}
    S_h(w) \sim \frac{w^{h_1 - 3/4} (1-w)^{h_2 - 3/4}}
    {2 \pi}
    \frac{2^{2h}}
    {h^{h_1 + h_2 - h_3 + h_4 - 1}}
    \cos\zeta(w, h)
    \qquad \text{as} ~ h \to \infty,
    \label{eq:S:asymptotics}
\end{equation}
where $\zeta$ is a phase defined by the equation
\begin{equation}
    e^{i \zeta(w, h)} = e^{i \pi (h_1 - 1/4)} \left( 1 - 2w - 2i \sqrt{w(1-w)} \right)^{h-1/2}.
    \label{eq:zeta}
\end{equation}
This implies that $S_h$ oscillates in a complicated pattern that depends on $w$.
On the other hand,
\begin{equation}
    T_h(w) \sim \frac{w^{h_1 - h_3 - 1/2} (1-w)^{-h_2 - 3/4}}
    {2\pi}
    \frac{2^{2h}}{h^{h_1 + h_2 - h_3 + h_4 - 1}}
     \sin[ \pi (h_2 + h_4 - h)] \rho(w)^{h-1/2}
    \label{eq:T:asymptotics}
\end{equation}
where
\begin{equation}
    \rho(w) = \frac{w}{(1 + \sqrt{1-w})^2}.
    \label{eq:rho}
\end{equation}
$T_h$ oscillates as well at large $h$, but in a regular way determined by the sine. The crucial observation here is that $\rho(w) < 1$ for all $w < 1$. The power $\rho(w)^h$ provides therefore an exponential suppression at large $h$, for all values of $w$ except at the boundary point $w = 1$.

In fact, the asymptotics of the t-channel block are very similar to those of the Euclidean conformal blocks. For instance, the position-space conformal block with identical operators is given by
\begin{equation}
    g_h(z) = z^{h - 2h_\phi} {}_2F_1(h, h; 2h; z),
    \label{eq:g}
\end{equation}
and at large $h$ it obeys
\begin{equation}
    g_h(z) \sim \frac{1}{2 z^{2h_\phi - 1/2} (1-z)^{1/4}}
    2^{2h} \rho(z)^{h-1/2}
    \qquad \text{as} ~ h \to \infty.
    \label{eq:g:asymptotics}
\end{equation}
It is well known that the Euclidean conformal block expansion converges exponentially fast as long as the radial coordinate $\rho(z)$ is inside the unit disc, $| \rho(z) | < 1$, which corresponds to nearly all Euclidean configurations of 4 points~\cite{Hogervorst:2013sma}.
The difference between eqs.~\eqref{eq:T:asymptotics} and \eqref{eq:g:asymptotics} is merely a power of $h$: no matter the sign of the exponent of $h^{h_1 + h_2 - h_3 + h_4 - 1}$, the exponential suppression provided by $\rho(w)^h$ make the t-channel expansion converge.

In summary, we are dealing with an asymmetric equation in which the t-channel side is absolutely convergent, but the s-channel side only converges in the distributional sense.

\subsubsection*{Example: the Ising model}

To get an idea of what this means in practice, let us examine a known correlation function. We focus on the operator $\sigma$ in the critical Ising model. This is a scalar primary with scaling dimension $\Delta = \frac{1}{8}$, or equivalently conformal weights $(h, \hbar) = \left( \frac{1}{16}, \frac{1}{16} \right)$. 
$\sigma$ is also a Virasoro primary, and its OPE contains operators from two Verma modules: the vacuum module, which includes the energy-momentum tensor, and another module whose highest-weight representation is the scalar operator $\varepsilon$ with $(h, \hbar) = \left( \frac{1}{2}, \frac{1}{2} \right)$. 
The Virasoro blocks for $\sigma$ are known, so that the CFT data, namely conformal weight and OPE coefficients, can be determined by an expansion in global conformal blocks. This data is given in Table~\ref{tab:Ising:sigmaOPE} for all operators with spin up to 12.

\begin{figure}
    \centering
    \includegraphics[width=0.9\linewidth]{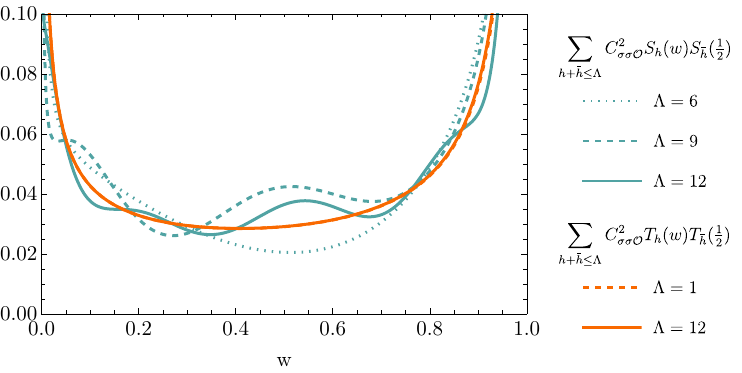}
    \caption{Comparison between the s- and t-channel expansions for the correlator $\langle \sigma \sigma \sigma \sigma \rangle$ in the Ising model, using the data from Table~\ref{tab:Ising:sigmaOPE}, as a function of $w$ with $\wbar = \frac{1}{2}$ fixed.}
    \label{fig:Ising-4sigma}
\end{figure}

\begin{table}
    \centering\bgroup
    \def\arraystretch{1.2}
    \setlength{\tabcolsep}{3.45mm}
    \begin{tabular}{|c||c|c|c|c|c|c|c|c|c|}
        \hline
        $\Delta$ & 1 & 2 & 4 & 4 & 5 & 6 & 6 & 7 & 8 \\
        \hline
        $\ell$ & 0 & 2 & 0 & 4 & 4 & 2 & 6 & 6 & 0 \\ 
        \hline
        $C_{\sigma\sigma\O}^2$ & $\frac{1}{4}$ & $\frac{1}{64}$ & $\frac{1}{4096}$ & $\frac{9}{40960}$ & $\frac{1}{65536}$ & $\frac{9}{2621440}$ & $\frac{25}{3670016}$ & $\frac{1}{1310720}$ & $\frac{81}{1677721600}$ \\
        \hline
    \end{tabular}
    
    \vspace{5mm}
    
    \setlength{\tabcolsep}{1.55mm}
    \begin{tabular}{|c||c|c|c|c|c|c|c|}
        \hline
        $\Delta$ & 8 & 8 & 9 & 9 & 10 & 10 & 10 \\
        \hline
        $\ell$ & 4 & 8 & 0 & 8 & 2 & 6 & 10 \\ 
        \hline
        $C_{\sigma\sigma\O}^2$ & $\frac{25}{234881024}$ & $\frac{15527}{57579405312}$ & $\frac{1}{1073741824}$ & $\frac{1125}{30064771072}$ & $\frac{45}{30064771072}$ & $\frac{15527}{3685081939968}$ & $\frac{251145}{20882130993152}$ \\
        \hline
    \end{tabular}
    
    \vspace{5mm}
    
    \setlength{\tabcolsep}{1.2mm}
    \begin{tabular}{|c||c|c|c|c|c|c|c|}
        \hline
        $\Delta$ & 11 & 11 & 12 & 12 & 12 & 12 \\
        \hline
        $\ell$ & 2 & 10 & 0 & 4 & 8 & 12 \\ 
        \hline
        $C_{\sigma\sigma\O}^2$ & $\frac{1}{21474836480}$ & $\frac{227}{120259084288}$ & $\frac{625}{13469017440256}$ & $\frac{46581}{786150813859840}$ & $\frac{251145}{1336456383561728}$ & $\frac{18598401}{32317945275219968}$ \\
        \hline
    \end{tabular}
    \egroup
    \caption{All primary operators with $\Delta \leq 12$ in the $\sigma \times \sigma$ OPE of the Ising model, togehter with the OPE coefficients squared. The first two entries in the top table correspond respectively to the scalar operator $\varepsilon$ and to the energy-momentum tensor $T$ with $C_{\sigma\sigma T} = 2h_\sigma = \frac{1}{8}$.
    Note that operators with non-zero spin enter twice in the OPE, with $(h, \hbar) = \left( \frac{\Delta \pm \ell}{2}, \frac{\Delta \mp \ell}{2} \right)$.}
    \label{tab:Ising:sigmaOPE}
\end{table}

Figure~\ref{fig:Ising-4sigma} shows the contribution of these operators to the crossing equation at a fixed value $\wbar = \frac{1}{2}$, while $w$ in allowed to vary freely in the interval $(0, 1)$. The convergence in the t channel is striking: there is barely any visible difference between the contribution of the first operator, the scalar $\varepsilon$ with $\Delta_\epsilon = 1$, and the sum involving all 39 operators with $\Delta \leq 12$. 
On the contrary, the s-channel convergence is quite slow: even though the three different truncations visible in the figure are clearly following the t-channel curve, it is evident that many more operators would be needed to obtain a reasonable match.

%%%%%%%%%%%%%%%%%%%%%%%%%%%%%%%%%%%%%%%%%%%%%%%%%%%
%%%%%%%%%%%%%%%%%%%%%%%%%%%%%%%%%%%%%%%%%%%%%%%%%%%

\section{Connection with position-space conformal blocks}
\label{sec:Euclideanconnection}

In this section, we answer the following important question: how is the momentum-space crossing equation related to the ordinary bootstrap equation in Euclidean position space?
Since the causality condition used as a starting point is specific to Minkowski space-time, it seems evident that the answer involves continuing the correlator at least all the way to the branch cuts found at the boundary of the Euclidean domain.
This is supported by the observation that the absolute convergence found on the first sheet is lost in the momentum-space equation.

However, since the Fourier transform involves integrating over all of Minkowski space-time, it is not at all obvious that a simple connection exists. Relating the Euclidean and Minkowski Fourier transforms is not helpful either: the procedure is already quite involved at the level of the 3-point function~\cite{Bautista:2019qxj, Gillioz:2021sce}, so it seems very difficult to achieve with a 4-point function.
But in fact, as we shall see now, a surprisingly simple connection can be established.

%%%%%%%%%%%%%%%%%%%%%%%%%%%%%%%%%%%%%%%%%%%%%%%%%%%

\subsection{A basis of Jacobi polynomials}

The distributional convergence of s-channel expansion suggests that we should integrate the crossing equation against some suitable functions in $w$ and $\wbar$. Since these only need to be functions defined over the unit interval, we can use polynomials.
There are many orthogonal bases of polynomials out there, but two stick out.
We study the first basis here and leave the second for Section~\ref{sec:newbasis} below.

The key observation is that the hypergeometric function in the s-channel block \eqref{eq:S} turns into a Jacobi polynomial of degree $n$ when $h = h_1 - h_2 + 1 + n$ with $n \in \mathbb{N}$. In the case of identical operators, this happens when $h$ is integer. We have
\begin{equation}
    S_h(w) = \frac{\Gamma(2h)}
    {\Gamma(h + h_1 + h_2 - 1) \Gamma(h + h_4 - h_3)}
    w^{2h_1 - 1} \widetilde{P}_n(w)
    \qquad
    \text{for}~
    h = h_1 - h_2 + 1 + n.
    \label{eq:Ptilde}
\end{equation}
Jacobi polynomials are characterized by two parameters that are related here to the conformal weights $h_1$ and $h_2$: in the standard notation, $\widetilde{P}_n(w) \equiv P_n^{(2h_1 - 1, 1 - 2h_2)}(1 - 2w)$.
They form a basis of functions on the unit interval, and satisfy an orthogonality relation given by
\begin{equation}
    \int_0^1 dw \, w^{2h_1 - 1} (1-w)^{1 - 2h_2}
    \widetilde{P}_j(w) \widetilde{P}_n(w)
    = \frac{\delta_{jn} \Gamma(2h_1 + n) \Gamma(2 - 2h_2 + n)}
    {n! (2h_1 - 2h_2 + 2n + 1) \Gamma(2h_1 - 2h_2 + n + 1)}.
\end{equation}
We can therefore define a family of functionals
\begin{equation}
    \widetilde{\omega}_j[f] \equiv \frac{(-1)^j j!}{\Gamma(2 - 2h_2 + j)}
    \int_0^1 dw (1-w)^{1-2h_2} \widetilde{P}_j(w) f(w),
    \label{eq:functional:omegatilde}
\end{equation}
such that the functional applied to $S_h$ has zeros at $h = h_1 - h_2 + 1 + n$ for every integer $n \neq j$. Using the orthogonality relation, we find
\begin{equation}
    \widetilde{\omega}_j[S_h] = \frac{(2 - 2h)_j}{\Gamma(h + h_4 - h_3)} \delta_{jn},
    \qquad\quad
    \text{for} ~ h = h_1 - h_2 + 1 + n.
\end{equation}
Remarkably, the integral of $S_h(w)$ and $T_h(w)$ can be solved analytically at arbitrary values of $h$, see Appendix~\ref{sec:integrals}. Defining
\begin{equation}
    \widetilde{S}_{h, j} \equiv \widetilde{\omega}_j[S_h],
    \qquad
    \widetilde{T}_{h, j} \equiv \widetilde{\omega}_j[T_h],
\end{equation}
we find the closed-form solutions
\begin{align}
    \widetilde{S}_{h, j}
    &= \frac{\Gamma(2h)}{\Gamma(h_2 - h_1 + h) \Gamma(h_4 - h_3 + h)}
    \frac{1}{\Gamma(h_1 - h_2 + h) \Gamma(h_1 - h_2 - h + 1)}
    \nonumber \\
    & \quad \times
    \frac{1}{(h_1 - h_2 + h + j) (h_1 - h_2 - h + 1 + j)},
    \label{eq:Sjtilde}
    \\
    \widetilde{T}_{h, j} &= \frac{(h_1 + h_3 - h)_j}
    {\Gamma(h_4 + h_2 - h) \Gamma(h_1 - 2h_2 - h_3 + h + j + 2)}
    \nonumber \\
    &\quad \times
    {}_4F_3\left( \begin{array}{c}
        h_1 - h_3 + h, h_4 - h_2 + h, h - h_1 - h_3 + 1, h - h_2 - h_4 + 1 \\
        2h, h_1 - 2h_2 - h_3 + h + j + 2, h - h_1 - h_3 - j + 1
    \end{array}; 1 \right)
    \label{eq:Tjtilde}.
\end{align}
In the case of identical external operators $h_i = h_\phi$, this becomes
\begin{align}
    \widetilde{S}_{h, j}
    &= \frac{\Gamma(2h)}{\Gamma(h)^2}
    \frac{\sin(\pi h)}{\pi (h + j) (1 - h + j)},
    \\
    \widetilde{T}_{h, j} &= \frac{\sin[\pi (2h_\phi - h)]}{\pi}
    \frac{(2h_\phi - h)_j}
    {(h - 2h_\phi + 1)_{j + 1}}
    {}_4F_3\left( \begin{array}{c}
        h, h, h - 2h_\phi + 1, h - 2h_\phi + 1 \\
        2h, h - 2h_\phi + j + 2, h - 2h_\phi - j + 1
    \end{array}; 1 \right).
\end{align}
The surprise is that this is precisely the basis of analytic functional derived in Ref.~\cite{Mazac:2016qev}.

%%%%%%%%%%%%%%%%%%%%%%%%%%%%%%%%%%%%%%%%%%%%%%%%%%%

\subsection{The position-space analytic functionals}

Let us quickly review the construction of these analytic functionals. For simplicity, we work with holomorphic operators so that the bootstrap equation is one-dimensional.
We also focus on identical operators,
the general case with four distinct operators being presented in Appendix~\ref{sec:Euclideanconnection:distinct}.

The 4-point correlation function of a holomorphic operator $\phi(z)$ in Euclidean position space admits a conformal block expansion of the form~\cite{Dolan:2000ut, Dolan:2003hv},
\begin{equation}
    \langle \phi(z_1) \phi(z_2) \phi(z_3) \phi(z_4) \rangle
    = \frac{1}{|z_1 - z_4|^{2h_\phi} |z_2 - z_3|^{2h_\phi}}
    \sum_\O C_{\phi\phi\O}^2 g_h(z),
\end{equation}
where $z$ is the conformal cross-ratio $z = (z_1 - z_2) (z_3 - z_4) / (z_1 - z_4) (z_3 - z_2)$,
and $g_h$ is the function defined above in eq.~\eqref{eq:g}, with $h$ the conformal weight of the operator $\O$.

The Euclidean correlator is symmetric under the exchange of any pair of points. In particular, the exchange $z_2 \leftrightarrow z_3$ corresponds to $z \leftrightarrow 1 - z$ and does not affect the factor multiplying the sum, so we must have the equality
\begin{equation}
    \sum_\O C_{\phi\phi\O}^2 g_h(z)
    = \sum_\O C_{\phi\phi\O}^2 g_h(1-z).
    \label{eq:crossing:z}
\end{equation}
This is the well-known conformal bootstrap equation in one dimension. It is valid for all $z$ in the interval $(0,1)$, and by analytic continuation in the entire complex $z$ plane with the exception of two branch cuts along $(-\infty, 0)$ and $(1, \infty)$. These cuts are found in every conformal block individually.

The idea of Ref.~\cite{Mazac:2016qev} is to study the behavior of the crossing equation along these branch cuts, and to project them on a basis of Legendre polynomials.
If we make the change of variable $z = 1 / (1 - x)$, the branch cut at $z \in (1, \infty)$ is mapped to the unit interval $x \in (0, 1)$. We define the discontinuity of the conformal block along this cut as
\begin{equation}
    f_h^+(x) \equiv \lim_{\varepsilon \to 0_+} \frac{1}{\pi}
    \im g_h\left( \frac{1}{1 - x} + i \varepsilon \right),
    \qquad
    \text{for} ~ x \in (0, 1).
    \label{eq:fplus:def}
\end{equation}
By symmetry, the same map computes the discontinuity of the opposite branch cut in the cross-channel, with $1 - z = x / (x - 1)$:
\begin{equation}
    f_h^-(x) \equiv \lim_{\varepsilon \to 0_+} \frac{1}{\pi}
    \im g_h\left( \frac{x}{x - 1} - i \varepsilon \right),
    \qquad
    \text{for} ~ x \in (0, 1).
    \label{eq:fminus:def}
\end{equation}
Using the explicit formula \eqref{eq:g} and some known identities for the hypergeometric function (see Appendix~\ref{sec:Euclideanconnection:distinct}), we can show that 
\begin{align}
    f_h^+(x) &= \frac{\Gamma(2h)}{\Gamma(h)^2} (1-x)^{2h_\phi}
    {}_2F_1(1-h, h; 1; x),
    \\
    f_h^-(x) &= \frac{\sin[\pi (2h_\phi - h)]}{\pi} x^{h - 2h_\phi} (1-x)^{2h_\phi}
    {}_2F_1(h, h; 2h; x).
\end{align}
The crossing equation implies that 
\begin{equation}
    \sum_\O C_{\phi\phi\O}^2 \left[ f_h^+(x) - f_h^-(x) \right] = 0.
\end{equation}
The next step is to apply a functional to this equation, defining
\begin{equation}
    \omega_j[f] \equiv (-1)^j \int_0^1 dx (1-x)^{-2h_\phi} P_j(1-2x) f(x),
\end{equation}
where the $P_j$ are Legendre polynomials, a special case of the Jacobi polynomials. 
Computing the integrals with the help of Appendix~\ref{sec:integrals}, we recover the functions found above in eqs.~\eqref{eq:Sjtilde} and \eqref{eq:Tjtilde},
\begin{equation}
    \omega_j[f_h^+] = \widetilde{S}_{h, j},
    \qquad\qquad
    \omega_j[f_h^-] = \widetilde{T}_{h, j},
\end{equation}
leading to a family of crossing equations
\begin{equation}
    \sum_\O C_{\phi\phi\O}^2 \left[ \widetilde{S}_{h,j} - \widetilde{T}_{h,j} \right] = 0
    \qquad\quad
    \forall ~ j = 0, 1, 2, \ldots
    \label{eq:omegafunctionalequation}
\end{equation}
This basis of equations has had a remarkable impact in one- and two-dimensional conformal field theory, as it could be used to build a variety of useful bootstrap functionals~\cite{Mazac:2016qev, Mazac:2018mdx, Mazac:2018ycv, Paulos:2019gtx, Mazac:2019shk, Paulos:2021jxx, Ghosh:2021ruh}, derive OPE inversion formulae and dispersion relations~\cite{Mazac:2018qmi, Caron-Huot:2020adz, Paulos:2020zxx}, and even establish surprising connections with pure mathematics~\cite{Hartman:2019pcd}.
For us, the fact that eq.~\eqref{eq:omegafunctionalequation} arises from the momentum-space crossing equation gives a non-trivial check of its validity.

%%%%%%%%%%%%%%%%%%%%%%%%%%%%%%%%%%%%%%%%%%%%%%%%%%%

\subsection{Integral transform of the Euclidean discontinuity}

We can actually go further than verifying the momentum-space crossing equation and directly link it to the position-space bootstrap along the cut.
To do so, note that the functional \eqref{eq:functional:omegatilde} can be inverted: since Jacobi polynomials form a basis of functions on the interval $(0, 1)$, we can write
\begin{equation}
    S_h(w) = w^{2h_\phi - 1} 
    \sum_{j = 0}^\infty \frac{(-1)^j (2j+1) j!}{\Gamma(2h_\phi + j)}
    \widetilde{S}_{h,j} \widetilde{P}_j(w).
\end{equation}
A similar expression hold for $T_h(w)$ expressed as an infinite sum of $\widetilde{T}_{h, j}$.

Using the fact that $\widetilde{S}_{h, j}$ is itself the integral transform of $f_h^+(x)$,
we can express $S_h(w)$ as the integral of the Euclidean block along the branch cut $z \in (1, \infty)$,  
\begin{equation}
    S_h(w) = \int_0^1 dx \mathcal{K}(w, x) f_h^+(x),
\end{equation}
where the integration kernel $\mathcal{K}(w, x)$ is defined by the infinite sum
\begin{equation}
    \mathcal{K}(w, x) = \frac{w^{2h_\phi - 1}}{(1-x)^{2h_\phi}} 
    \sum_{j = 0}^\infty \frac{(2j+1) j!}{\Gamma(2h_\phi + j)}
    \widetilde{P}_j(w) P_j(1-2x).
    \label{eq:K}
\end{equation}
Similarly, the t-channel conformal block correspond to the same integral transform applied to the opposite branch cut of the position-space block at $z \in (-\infty, 0)$,
\begin{equation}
    T_h(w) = \int_0^1 dx \mathcal{K}(w, x) f_h^-(x).
\end{equation}

A closed-form expression for the integration kernel $\mathcal{K}(w, x)$ is not known to us.
Nevertheless, it can be shown that the sum in eq.~\eqref{eq:K} is absolutely convergent in the open interval $w, x \in (0,1)$ when $h_\phi > 1$: since Jacobi polynomials (including Legendre polynomials as a special case) decay as $j^{-1/2}$ at large $j$, the summand asymptotes to $j^{1-2h_\phi}$.
At and below the critical value $h_\phi = 1$, it is possible to define the sum by analytic continuation in $h_\phi$, for all $h_\phi > 0$. 
A special case is encountered at $h_\phi = \frac{1}{2}$: the Jacobi polynomial $\widetilde{P}_j$ is itself a Legendre polynomial and the kernel becomes
\begin{equation}
    \mathcal{K}(w, x) = \frac{1}{1-x}
    \sum_{j = 0}^\infty (2j+1) P_j(1-2w) P_j(1-2x)
    = \frac{\delta(w -x)}{1-x}.
\end{equation}
Note that this is a formal equality: the sum does not converge at any $x$ and $w$, but the equality is true in a distributional sense since Legendre polynomials form an orthogonal basis of polynomials on the unit interval in $w$ and in $x$.
In this special case, the momentum-space conformal blocks are actually directly related to the discontinuity of the position-space blocks,
\begin{equation}
    S_h(w) = \frac{f_h^+(w)}{1-w},
    \qquad
    T_h(w) = \frac{f_h^-(w)}{1-w},
    \qquad
    \text{for}~
    h_\phi = \frac{1}{2}.
\end{equation}

%%%%%%%%%%%%%%%%%%%%%%%%%%%%%%%%%%%%%%%%%%%%%%%%%%%
%%%%%%%%%%%%%%%%%%%%%%%%%%%%%%%%%%%%%%%%%%%%%%%%%%%

\section{A basis of analytic functionals}
\label{sec:newbasis}

The family of functionals $\widetilde{\omega}_j$ is nice because it has zeros at interesting values of the internal conformal weight: at integer $h$ in the s channel, and at the double-twist values $h = 2 h_\phi + n$ in the t channel. 
When the two sets of zeros are aligned, which is the case at (half-)integer $h_\phi$, this has been used to construct optimal functionals~\cite{Mazac:2016qev}. The generalization to other values of $h_\phi$ is however more involved~\cite{Mazac:2018mdx}. A basis of functionals with matching zeros in the s and t channels was constructed as contour integral in cross-ratio space in Ref.~\cite{Mazac:2019shk}. In this section, we show that the same functionals arise naturally from the momentum-space crossing equation. We provide a new closed-form expression for these functionals and a generalization to the case of distinct operators.

%%%%%%%%%%%%%%%%%%%%%%%%%%%%%%%%%%%%%%%%%%%%%%%%%%%

\subsection{Jacobi polynomials again}

The starting point for constructing the functionals $\widetilde{\omega}_j$ in Section~\ref{sec:Euclideanconnection} was the observation that the conformal block $S_h(w)$ is polynomial at integer values of $h$.
In fact, these are not the only special values: a similar behavior is found when $h = h_1 + h_2 + n$ with $n \in \mathbb{N}$.
Using the identity ${}_2F_1(a, b; c; w) = (1-w)^{c-a-b} {}_2F_1(c-a, c-b; c; w)$, we can rewrite the s-channel block as
\begin{equation}
    S_h(w) = \frac{\Gamma(2h) w^{2h_1 - 1} (1-w)^{2h_2 - 1}}
    {\Gamma(h + h_4 - h_3) \Gamma(h + h_2 - h_1) \Gamma(2 h_1)}
    {}_2F_1( h_1 + h_2 + h - 1, h_1 + h_2 - h; 2 h_1; w).
    \label{eq:S:alt}
\end{equation} 
When $h = h_1 + h_2 + n$ with $n \in \mathbb{N}$, one of the parameters of the ${}_2F_1$ function is a negative integer, so the infinite hypergeometric series terminates. The result is again proportional to a Jacobi polynomial of degree $n$ in $w$,
\begin{equation}
    S_h(w) = \frac{\Gamma(2h) \Gamma(h - h_1 - h_2 + 1) w^{2h_1 - 1} (1-w)^{2h_2 - 1}}
    {\Gamma(h + h_1 - h_2) \Gamma(h + h_2 - h_1) \Gamma(h + h_4 - h_3)}
    \widehat{P}_n(w)
    \qquad
    \text{for}~
    h = h_1 + h_2 + n.
    \label{eq:hat}
\end{equation}
In the standard notation, $\widehat{P}_n(w) = P_n^{(2h_1 - 1, 2h_2 - 1)}(1 - 2w)$. Note that the parameters of this polynomial, $2h_2 - 1$ and $2h_2 - 1$, are both larger than the critical value $-1$ in unitary theories. This means that the $\widehat{P}_n(w)$ form an actual orthogonal basis of functions in the unit interval $w \in (0,1)$, whereas their cousins $\widetilde{P}_n(w)$ of Section~\ref{sec:Euclideanconnection} require working by analytic continuation in some cases.
The orthogonality relation is given by
\begin{equation}
    \int_0^1 dw \, w^{2h_1 - 1} (1-w)^{2h_2 - 1}
    \widehat{P}_j(w) \widehat{P}_n(w)
    = \frac{\delta_{jn} \Gamma(2h_1 + n) \Gamma(2h_2 + n)}
    {n! (2h_1 + 2h_2 + 2n - 1) \Gamma(2h_1 + 2h_2 + n - 1)}.
\end{equation}
This suggests defining again a family of functionals
\begin{equation}
    \widehat{\omega}_j[f] \equiv \frac{(-1)^j j!}{\Gamma(2h_2 + j)}
    \int_0^1 dw \widehat{P}_j(w) f(w),
    \label{eq:functional:omegahat}
\end{equation}
such that $\widehat{\omega}_j$ applied to $S_h$ has zeros at every $h = h_1 + h_2 + n$ with $n \neq j$:
\begin{equation}
    \widehat{\omega}_j[S_h] = \frac{j! (2 - 2h)_j}
    {\Gamma(h + h_2 - h_1) \Gamma(h + h_4 - h_3)} \delta_{jn}
    \qquad\quad
    \text{for} ~ h = h_1 + h_2 + n.
    \label{eq:Sj:integer}
\end{equation}
For convenience, we introduce the notation
\begin{equation}
    S_{h, j} \equiv \widehat{\omega}_j[S_h],
    \qquad
    T_{h, j} \equiv \widehat{\omega}_j[T_h].
\end{equation}
The results of Appendix~\ref{sec:integrals} let us once again compute $S_h$ and $T_h$ in closed form:
\begin{align}
    S_{h, j}
    &= \frac{\Gamma(2h)}{\Gamma(h_2 - h_1 + h) \Gamma(h_4 - h_3 + h)}
    \frac{1}{\Gamma(h_1 + h_2 + h - 1) \Gamma(h_1 + h_2 - h)}
    \nonumber \\
    & \quad \times
    \frac{1}{(h_1 + h_2 - h + j) (h_1 + h_2 + h - 1 + j)},
    \label{eq:Sj}
    \\
    T_{h, j} &= \frac{(h - h_1 - h_3 - j + 1)_j}
    {\Gamma(h_4 + h_2 - h) \Gamma(h_1 + 2h_2 - h_3 + h + j)}
    \nonumber \\
    &\quad \times
    {}_4F_3\left( \begin{array}{c}
        h + h_1 - h_3, h + h_2 - h_4, h - h_1 - h_3 + 1, h + h_2 + h_4 - 1 \\
        2h, h_1 + 2h_2 - h_3 + h + j, h - h_1 - h_3 - j + 1
    \end{array}; 1 \right)
    \label{eq:Tj}.
\end{align}
Even though $S_{h, j}$ and $T_{h,j}$ appear to be quite similar to the $\widetilde{S}_{h,j}$ and $\widetilde{T}_{h,j}$ of eqs.~\eqref{eq:Sjtilde} and \eqref{eq:Tjtilde}, they are actually nicer: the zeros of the s and t channels are now aligned, both at the dimensions of composite operators $[\O_1 \partial^n \O_2]$.

In the special case $h_2 = \frac{1}{2}$, there is no distinction between the two functionals \eqref{eq:functional:omegahat} and \eqref{eq:functional:omegatilde}: $S_{h,j} = \widetilde{S}_{h,j}$ and $T_{h,j} = \widetilde{T}_{h,j}$ are actually respectively equal to the Euclidean discontinuities \eqref{eq:discontinuity:s} and \eqref{eq:discontinuity:t} of the position-space conformal blocks.

When all external operators are identical, $h_i = h_\phi$ for $i = 1, 2, 3, 4$, this reduces to the result quoted in Eqs.~\eqref{eq:Sj:identical} and \eqref{eq:Tj:identical} in the introduction.  In this case, the functionals $S_{h,j}$ and $T_{h,j}$ correspond to integrals of the position-space conformal block $g_h(z)$ given in eq.~\eqref{eq:g} along a complex contour,
\begin{equation}
    S_{h, j} = \int\limits_{\frac{1}{2} - i \infty}^{\frac{1}{2} + i \infty}
    \frac{dz}{2\pi i} H_j(z) g_h(z),
    \qquad\qquad
    T_{h, j} = \int\limits_{\frac{1}{2} - i \infty}^{\frac{1}{2} + i \infty}
    \frac{dz}{2\pi i} H_j(1-z) g_h(z)
\end{equation}
with an integration kernel found in Ref.~\cite{Mazac:2019shk} (with a different normalization)
\begin{equation}
    H_j(z) = \frac{1}{\Gamma(2h_\phi)^2} z^{-1}
    {}_3F_2(1, -j, 4h_\phi + j - 1; 2h_\phi, 2h_\phi; z^{-1}).
\end{equation}

In spite of its complicated form, $T_{h,j}$ is an analytic function of $h$ over the whole domain $h \geq 0$. For instance, eq.~\eqref{eq:Tj:identical} can be written as the infinite sum
\begin{equation}
    T_{h, j} = 
    \frac{1}{\Gamma(2h_\phi + h - 1) \Gamma(2h_\phi - h)}
    \sum_{k = 0}^\infty
    \frac{(h)_k^2 (h - 2h_\phi - j + k + 1)_j}
    {k! \Gamma(2h)_k (2h_\phi + h + k - 1)_{j + 1}}
\end{equation}
Each term in this sum is obviously analytic, and the sum is absolutely convergent since the summand decays like $k^{-2}$ at large $k$.

\subsubsection*{The identity operator}

The contribution of the identity operator can be determined taking the limit $h \to 0_+$, after setting $h_1 = h_2$ and $h_3 = h_4$ in the s channel, and $h_1 = h_3$ and $h_2 = h_4$ in the t channel. We find
\begin{equation}
    S_{0, j} = 0,
    \qquad\qquad
    T_{0, j} = \frac{(-1)^j (2h_1)_j}{\Gamma(2h_2)^2 (2h_2)_j}.
    \label{eq:T0j}
\end{equation}
In the case $h_1 = h_2 = h_\phi$, this gives eq.~\eqref{eq:T0j:identical}.
This limit agrees with the action of the functional $\widehat{\omega}_j$ on the identity blocks $S_0(w)$, which is identically zero, and $T_0(w)$ given by eq.~\eqref{eq:T:0}. Note however that integrating the distribution $\delta(w)$ requires extending the integral in eq.~\eqref{eq:functional:omegahat} beyond the interval $w \in (0, 1)$.

\subsubsection*{Convergence}

This observation raises an issue regarding the convergence of the OPE: the case can be made for the distributional convergence of the momentum-space crossing equation, since after all the kinematic variables $w$ and $\wbar$ are directly related to physical momenta of the Wightman correlation function. However, Jacobi polynomials are clearly not good test functions: they do not decay exponentially as $|w| \to \infty$, quite the opposite.
For this reason, it is legitimate to question the convergence of the OPE in terms of $S_{h,j}$ and $T_{h,j}$.

In fact, there is substantial evidence that the functionals $\widehat{\omega}_j$ cannot be commuted with the infinite sum of the OPE: applying any of them to the example of the correlators $\langle \sigma\sigma\sigma\sigma \rangle$ in the Ising model, as in Figure~\ref{fig:Ising-4sigma}, we do not observe a convergent behavior. 
The situation is quite similar in this respect to the functionals $\widetilde{\omega}_j$ in Section~\ref{sec:Euclideanconnection}, which were found not to be individually ``swappable'' with the OPE, but only in some linear combinations~\cite{Qiao:2017lkv}. 
Even though the question of the OPE convergence is crucial in the context of conformal bootstrap applications, we postpone it to further studies. As we shall see below, a number of consistent results are found that give us confidence in the value of these functionals.

\begin{figure}
    \centering
    \includegraphics[width=0.85\linewidth]{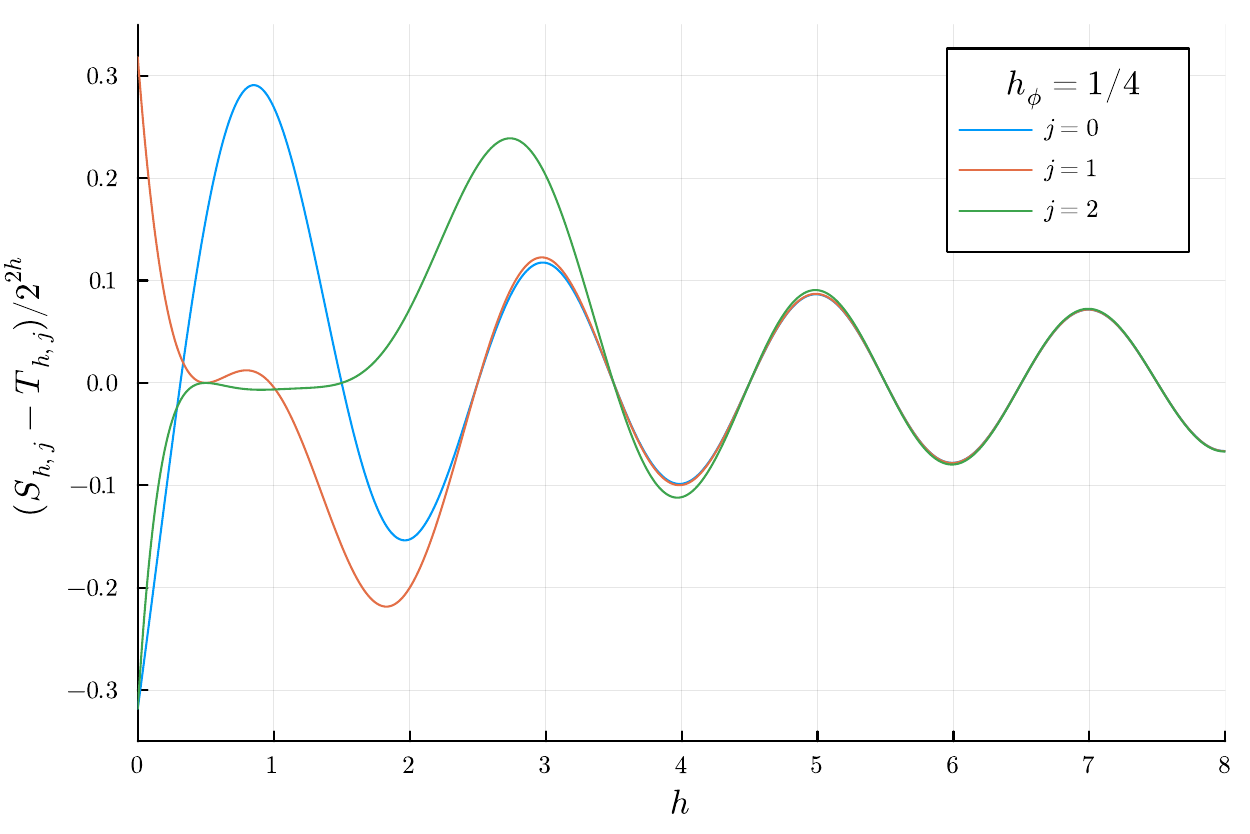}
    \caption{Difference between the s- and t-channel holomorphic blocks as a function of $h$, after applying the functional $\widehat{\omega}_j$ with $j = 0, 1, 2$,
    in the case of 4 identical external operators with $h_\phi = \frac{1}{4}$.
    The functionals have zeros at every $h = 2h_\phi + n$ with integer $n \neq j$
    (in this case half-integer values of $h$).}
    \label{fig:SminusT}
\end{figure}

One aspect that can be easily studied is the asymptotic behavior of the blocks as $h \to \infty$. For $S_{h, j}$, it is simple to check that
\begin{equation}
    S_{h, j}
    \sim \frac{1}{2 \pi^{3/2}}
    \frac{2^{2h}}{h^{h_1 + 3h_2 - h_3 + h_4 - 1/2}}
    \sin\left[ \pi (h - h_1 - h_2) \right]
    \qquad
    \text{for} ~ h \to \infty,
\end{equation}
independently of $j$. $S_{h, j}$ has the same type of asymptotic behavior as $S_h(w)$, see eq.~\eqref{eq:S:asymptotics}, but this time the oscillatory factor is periodic and the power of $h$ is different.
In the t channel, the exponential suppression found for $w < 1$ is lost when integrating over $w$ in the interval $(0, 1)$: the dominant contribution comes from the regime $w \sim 1$, and we find the remarkable relation
\begin{equation}
    \frac{T_{h, j}}{S_{h, j}} \sim - \frac{\sin\left[ \pi (h - h_1 - h_2) \right]}
    {\sin\left[ \pi (h - h_2 - h_4) \right]}
    \qquad
    \text{for} ~ h \to \infty.
\end{equation}
The phases cancel in the case of identical operators, so that
\begin{equation}
    T_{h, j} \sim -S_{h,j}
    \qquad
    \text{for} ~ h \to \infty, ~ h_i = h_\phi.
\end{equation}
Figure~\ref{fig:SminusT} shows the difference $S_{h,j} - T_{h,j}$ for a specific choice $h_\phi = \frac{1}{4}$ and several values of $j$. The difference quickly approaches the asymptotic form $S_{h, j} - T_{h, j} \sim 2S_{h, j} \sim -\cos(\pi h) 2^{2h} /(\pi^{3/2} \sqrt{h})$ at large $h$.

Note that in the case of identical external operators we have another remarkable relation at large $j$ and fixed $h$,
\begin{equation}
    T_{h, j} \sim S_{h,j}
    \qquad
    \text{for} ~ j \to \infty, ~ h_i = h_\phi.
\end{equation}
It implies that operators of small conformal weight $h$ decouple from the crossing equation after applying the functional $\widehat{\omega}_j$ with $j \gg 1$.

%%%%%%%%%%%%%%%%%%%%%%%%%%%%%%%%%%%%%%%%%%%%%%%%%%%

\subsection{Generalized free field theories}

Generalized free field (GFF) theories have OPEs that only involve composite operators. Since  the $S_{h, j}$ and $T_{h, j}$ have zeros precisely at these values of the conformal weights, they are particularly well-suited to study crossing in such theories.

By definition, only correlation functions of pairwise-identical operators exist in GFF. The contribution of the identity in one channel must be matched by an infinite tower of composite operators in the other.
If $\O_1 = \O_2$ and $\O_3 = \O_4$, the identity is in the s channel and the composite operators in the t channel.
Both give identically vanishing contributions to the crossing equation.

The interesting situation is $\O_1 = \O_3$ and $\O_2 = \O_4$: the composite operators of the form $[\O_1 \partial_+^n \partial_-^{\bar{n}} \O_2]$ that appear in the s channel must match the identity in the t channel, leading to a family of equations
\begin{equation}
    \sum_{n, \bar{n} = 0}^\infty \left| C_{n, \bar{n}} \right|^2 S_{h, j} S_{\hbar, \jbar}
    = T_{0, j} T_{0, \jbar},
    \qquad
    \forall ~ j, \jbar = 0, 1, 2, \ldots
\end{equation}
where $h = h_1 + h_2 + n$ and $\hbar = \hbar_1 + \hbar_2 + \bar{n}$, and $C_{n, \bar{n}}$ is the OPE coefficient of the corresponding operator.

GFF is special because it can be fully factorized into holomorphic and anti-holomorphic parts: this is true at the level of the correlation function, but also for each conformal block. This means that we can write $C_{n, \bar{n}} = C_n C_{\bar{n}}$
and focus on the one-dimensional crossing equation
\begin{equation}
    \sum_{n = 0}^\infty \left| C_n \right|^2 S_{h,j}
    = T_{0, j}
    \qquad
    \forall ~ j = 0, 1, 2, \ldots.
\end{equation}
Since $S_{h, j} \propto \delta_{nj}$, the action of the functional $\widehat{\omega}_j$  selects a single operator in the entire spectrum.
We must have
\begin{equation}
    | C_n |^2
    = \frac{T_{0,n}}{S_{h_1 + h_2 + n, n}}.
\end{equation}
Using eq.~\eqref{eq:Sj:integer} for the value of $S_{h, j}$ and eq.~\eqref{eq:T0j} for $T_{0,j}$, we obtain
\begin{equation}
     |C_n|^2 = \frac{(2h_1)_n (2h_2)_n}{n! (2h_1 + 2h_2 + n - 1)_j}.
\end{equation}
Note that the right-hand side is strictly positive for $h_1, h_2 > 0$.
In the limit of identical operators $h_1, h_2 \to h_\phi$, we must discard the operators with odd $n$, and we get a factor of 2 for even-spin operators, in agreement with eq.~\eqref{eq:identicallimit}. We recover the OPE coefficient of the generalized free boson, see for instance Ref.~\cite{Paulos:2019fkw},
\begin{equation}
     C_{\phi\phi\O}^2 = \frac{2\Gamma(h)^2 \Gamma(h + 2h_\phi - 1)}{\Gamma(2h_\phi)^2 \Gamma(2h - 1) \Gamma(h - 2h_\phi + 1)},
     \label{eq:GFFcoefficients}
\end{equation}
where $h = 2h_\phi + 2n$.
The odd-$n$ coefficients are not unphysical: they correspond to the free fermion OPE, obtained in the situation of anticommuting operators.
Eq.~\eqref{eq:GFFcoefficients} is still valid in this case, but with $h = 2h_\phi + 2n + 1$.

%%%%%%%%%%%%%%%%%%%%%%%%%%%%%%%%%%%%%%%%%%%%%%%%%%%

\subsection{One-dimensional bootstrap functionals}
\label{sec:bootstrapfunctionals}

To illustrate further the potential of the momentum-space conformal bootstrap, we construct a few examples of useful functionals in the case of identical, holomorphic operators.
Using the basis provided by $\widehat{\omega}_j$, we write
\begin{equation}
    \gamma(h) = \Gamma(2h_\phi)^2 \sum_{j = 0}^\infty (-1)^j
    a_j (S_{h, j} - T_{h, j}),
    \label{eq:functional:gamma}
\end{equation}
where the $a_j$ are real coefficients.
These functionals have a nice properties: at $h = 2 h_\phi + n$ with $n \in \mathbb{N}$, all conformal blocks vanish except for $S_{h, n}$, so we have 
\begin{equation}
    \gamma(2h_\phi + n) = \frac{n! (4h_\phi - 1 + n)_n} {(2h_\phi)_n^2} \, a_n.
\end{equation}
Note that the coefficients multiplying $a_n$ are all strictly positive when $h_\phi > 0$.
If we want to construct optimal functionals in the bootstrap sense, that is with positivity above a certain value $h = h_*$, then we must take $a_j > 0$ for all $j > h_* - 2 h_\phi$.
This is a necessary condition, but it is not sufficient: $S_{h,j} - T_{h,j}$ oscillates with period 2 in $h$, so $\gamma(h)$ can in general have zeros between the values $h = 2h\phi + n$. 

At the same time, the functional \eqref{eq:functional:gamma} is such that the contribution of the identity operator is
\begin{equation}
    \gamma(0) = -\sum_{j = 0}^\infty a_j.
    \label{eq:functional:gamma:0}
\end{equation}
If the $a_j$ are all positive, then the identity contribution is negative and there are always infinitely many solutions to crossing. 
Optimal functionals that annihilate the identity block are therefore constructed with $a_j > 0$ for all $j > 0$, and $a_0 < 0$, paying attention to the convergence: the $a_j$ must decay sufficiently fast at large $j$ for the sum \eqref{eq:functional:gamma:0} to converge. We provide a numerical construction of such functionals below.

Another interesting family of functionals is obtained choosing the $a_j$ so that the zeros of $\gamma$ are precisely aligned with the spectrum of a theory, assuming that it is known.
In this case, it is possible to isolate the contribution of given operator in the spectrum, and to compare its contribution to that of the identity, hence determining the OPE coefficient.
The simplest realization of this idea is the functional $\widehat{\omega}_j$ itself, corresponding to the case of a single non-zero $a_j$. But we shall see below that other functionals of this type can be constructed numerically.

\subsubsection*{Optimal functionals for generalized free fermion theories}

The construction of optimal functionals is the central problem of the numerical conformal bootstrap: if one can find a functional that annihilates the identity operator and that is non-negative for all values of $h$ above a certain $h_*$, then $h_*$ gives an upper bound on the lowest-lying operator in the spectrum.
Moreover, if a theory saturates this bound, its spectrum is determined by the zeros of the optimal functional. Since the functional is non-negative, these zeros must be (at least) double zeros.
In general, we expect therefore that optimal functionals have double zeros at all values of $h$ corresponding to operators of the theory.

In the case of generalized free field theory such as the free fermion introduced earlier, these double zeros must appear at $h = 2h_\phi + 2n + 1$ for $n \in \mathbb{N}$.
The $\widehat{\omega}_j$ are a good basis for constructing such an optimal functional: indeed, a functional in the form of eq.~\eqref{eq:functional:gamma} with only even $j$ in the sum has already simple zeros at these values. The problem reduces therefore to choosing the $a_{2j}$ such that the zeros at $h = 2h_\phi + 2n + 1$ are actually double zeros. 
This is still a difficult problem.
In the case $h_\phi = \frac{1}{2}$, corresponding to the free fermion theory, it was solved in Ref.~\cite{Mazac:2016qev}. Since the $S_{h, j}$ and $T_{h, j}$ coincide with the basis used in that case, the solution found there is valid here. We can take 
\begin{equation}
    a_j = (2j+1) \left[ \frac{1}{(j+1)(j+2)} - \left( j (j+1) + \frac{1}{2} \right)
    \Psi'\left( \frac{j+1}{2} \right) - 2 \right]
    \quad
    \text{for} ~ j ~ \text{even},
\end{equation}
where $\Psi(x) = \psi(x + 1/2) - \psi(x)$ in terms of the digamma function $\psi$.

Beyond this special case, we do not know how to construct the optimal functionals analytically, but an efficient numerical approach can be formulated as follows.
We first extract from the conformal blocks a factor that depends on $h$ but is independent of $j$, and define
\begin{equation}
    G_{h, j} = \frac{\Gamma(h)^2 \Gamma(2h_\phi - h) \Gamma(h + 2h_\phi - 1)}{\Gamma(2h)}
    \left[ S_{h,j} - T_{h,j} \right].
\end{equation}
$G_{h, j}$ does not have the same zeros as $S_{h, j}$ and $T_{h, j}$. It is analytic in $h$ apart from a pole at $h = 2h_\phi + j$. Explicitly,
\begin{equation}
    G_{h, j} = \frac{1}{(2h_\phi - h + j)(2h_\phi + h - 1 + j)}
    - \sum_{k = 0}^\infty
    \frac{(k + 1)_{h - 1} (h - 2h_\phi - j + k + 1)_j}
    {(h + k)_h \Gamma(h + 2h_\phi + k - 1)_{j + 1}}.
\end{equation}
The sum over $k$ is absolutely convergent and can be evaluated efficiently: each term satisfies recursion relations in $h$ and $j$ that involve rational functions of the parameters.
\begin{figure}
    \centering
    \includegraphics[width=0.85\linewidth]{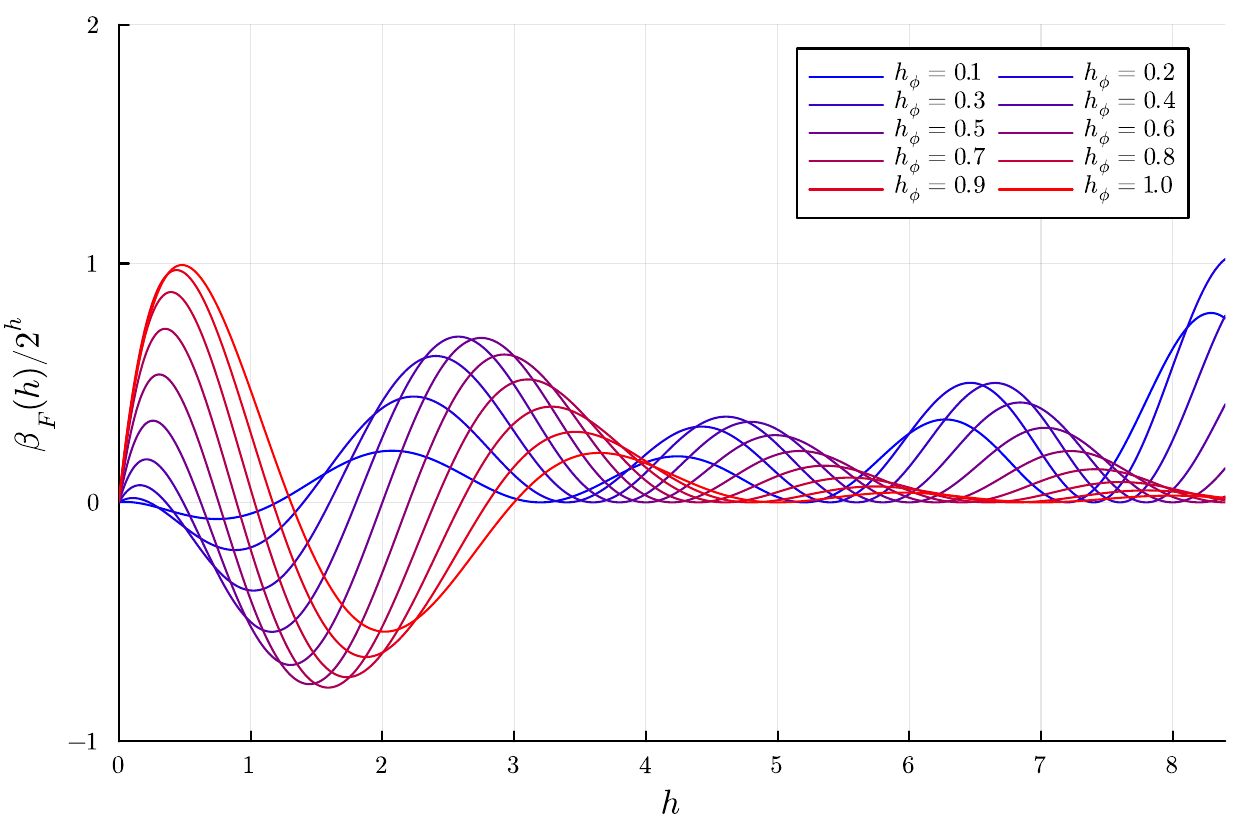} 
    \caption{Optimal functionals annihilating the spectrum of the generalized free fermion theories, shown for a variety of conformal weights $h_\phi$ of the external operator.
    The curves have double zeros at $2h_\phi + 2n + 1$ with $n = 1, 2, 3, \ldots$ and a simple zero at $h = 2h_\phi + 1$.}
    \label{fig:betafunctionals}
\end{figure}

Let us now define, for a certain integer $\Lambda$, a matrix of size $\Lambda \times \Lambda$,
\begin{equation}
    \mathcal{B}_{nj} = G_{2h_\phi + 2n + 1, 2j}
    \qquad
    \text{for} ~ n, j = 0, 1, 2, \ldots, \Lambda - 1,
\end{equation}
and consider the matrix equation
\begin{equation}
    \sum_{j = 0}^{\Lambda - 1} \mathcal{B}_{nj} a_{2j} = \delta_{n0}
    \qquad
    \text{for} ~ n = 0, 1, 2, \ldots, \Lambda - 1.
\end{equation}
The $a_j$ that solve this equation define a functional with double zeros at all values of $h = 2h_\phi + 2k + 1$ with $k = 1, 2, 3, \ldots, \Lambda - 1$, and a simple zero at $h = 2h_\phi + 1$:
\begin{equation}
    \beta^\Lambda(h) = \Gamma(2h_\phi)^2 \sum_{j = 0}^{\Lambda - 1}
    \mathcal{B}_{j0}^{-1} (S_{h, 2j} - T_{h, 2j}).
    \label{eq:functional:betaB:truncated}
\end{equation}
The inverse $\mathcal{B}_{jn}^{-1}$ of $\mathcal{B}_{nj}$ can be computed numerically, and the entries of its first column at $n = 0$ converge rapidly in the limit $\Lambda \gg 1$. We observe that the value $\beta^\Lambda(0)$ corresponding to the contribution of the identity operator approaches zero rapidly as $\Lambda \gg 1$, even though we did not impose this as a constraint. We conclude that the functional 
\begin{equation}
    \beta(h) = \lim_{\Lambda \to \infty} \beta^\Lambda(h)
    \label{eq:functional:betaB}
\end{equation}
is the optimal functional saturated by the generalized free fermion theory.
Figure~\ref{fig:betafunctionals} shows the numerical functionals computed with $\Lambda = 25$ for a variety of values of $h_\phi$.
These numerical results agree with the integral representation for the generalized free fermion functionals given in Ref.~\cite{Mazac:2018mdx}.

\subsubsection*{Numerical functionals with simple zeros at integer $h$}

Another interesting type of functional can be built from the basis of $S_{h,j}$ and $T_{h,j}$. It is motivated by an observation made in Appendix~\ref{sec:4statePotts}: the momentum-space crossing equation takes a simple form when applied to the 4-point function of the energy operator $\varepsilon$ in the 4-state Potts model~\cite{Gorbenko:2018ncu, Gorbenko:2018dtm}. The operator $\varepsilon$ has conformal weights $(h, \hbar) = \left(\frac{1}{4}, \frac{1}{4}\right)$ and the particularity that only the vacuum module enter its OPE. In other words, it obeys the fusion rule
\begin{equation}
    \varepsilon \times \varepsilon = 1.
\end{equation}
It is therefore natural to ask whether one can build functionals in the form of eq.~\eqref{eq:functional:gamma} that have zeros at all but one integer values of $h$. If such functionals can be found, then the OPE coefficients can be computed matching the contribution of the corresponding operator with that of the identity.

The question can obviously be generalized beyond the 4-state Potts model to external conformal weights $h_\phi \neq \frac{1}{4}$.
Assuming that there are only operators with integer conformal weights in the OPE $\phi \times \phi$, can we determine their coefficients?
\begin{figure}
    \centering
    \includegraphics[width=0.85\linewidth]{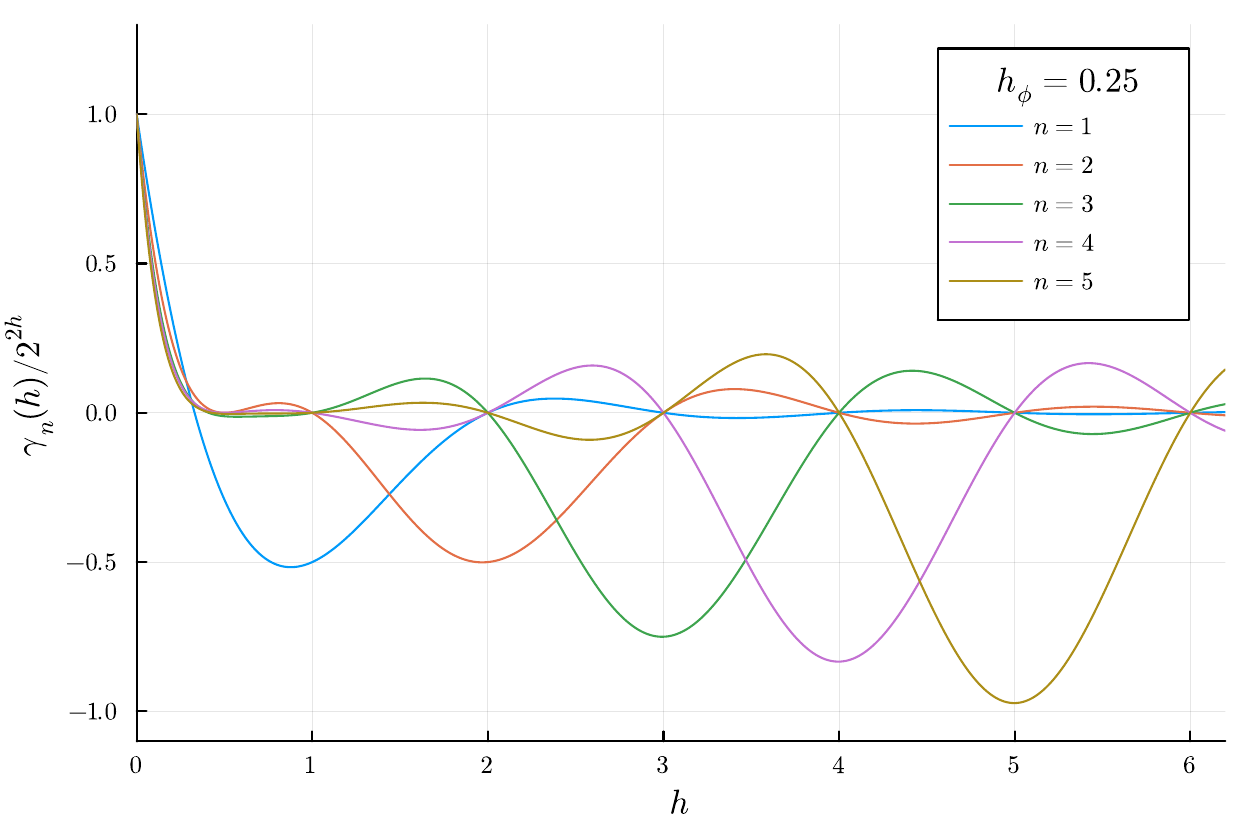}
    \caption{Functionals $\gamma_n$ applied to the one-dimensional crossing equation with $h_\phi = \frac{1}{4}$, displaying zeros at every integer $h \neq n$.
    The functionals are normalized so that $\gamma_n(0) = 1$.}
    \label{fig:gammafunctionals}
\end{figure}

We proceed as for the general free fermion functional by defining a square matrix whose entries correspond to $G_{h, j}$ evaluated at the values of $h$ at which the functional should vanish,
\begin{equation}
    \mathcal{A}_{kj} = G_{k, j}
    \qquad
    \text{for} ~ k, j = 0, 1, 2, \ldots, \Lambda - 1.
\end{equation}
The notable difference in this case is that all values of $j$ appear, even and odd.
The matrix equations that we want to solve are of the form
\begin{equation}
    \sum_{j = 0}^{\Lambda - 1} \mathcal{A}_{kj} a_j = \delta_{nk},
    \qquad
    \text{for} ~ n, k = 0, 1, 2, \ldots, \Lambda - 1.
\end{equation}
There is one such equation for each $n$. The coefficients $a_j$ defining the functional $\gamma$ in eq.~\eqref{eq:functional:gamma} are found by inversion of the matrix $\mathcal{A}$: we define
\begin{equation}
    \gamma_n^\Lambda(h) = \Gamma(2h_\phi)^2 \sum_{j=0}^{\Lambda - 1}
    \mathcal{A}_{jn}^{-1} \left( S_{h, j} - T_{h, j} \right)
\end{equation}
and take the limit 
\begin{equation}
    \gamma_n(h) = \lim_{\Lambda \to \infty} \gamma_n^\Lambda(h).
\end{equation}
As before, the numerical convergence is very good. With $\Lambda = 25$ and $h_\phi = \frac{1}{4}$, as in the 4-state Potts model, we obtain the functionals shown in Fig.~\ref{fig:gammafunctionals}. We can see that the zeros are precisely aligned at integer $h$.

When applied to a theory whose spectrum is integer-valued, the functional $\gamma_n$ selects a single operator with $h = n$ in the OPE. The crossing equation reads
\begin{equation}
    \gamma_n(0) + C_n^2 \gamma_n(n) = 0,
\end{equation}
and the square of the OPE coefficient $C_n$ can therefore be computed from the ratio of $\gamma_n(0)$ and $\gamma_n(n)$. By construction, we have
\begin{equation}
    \gamma_n^\Lambda(n) = \Gamma(2h_\phi)^2
\end{equation}
and
\begin{equation}
    \gamma_n^\Lambda(0) = - \sum_{j=0}^{\Lambda - 1} (-1)^j \mathcal{A}_{jn}^{-1}.
\end{equation}
We see numerical convergence of this sum in the limit $\Lambda \to \infty$ for all $h_\phi < \frac{1}{2}$. Above this value the convergence is lost and we conclude that the $\gamma_n$ functionals cannot be constructed in this simple way.
\begin{figure}
    \centering
    \includegraphics[width=0.85\linewidth]{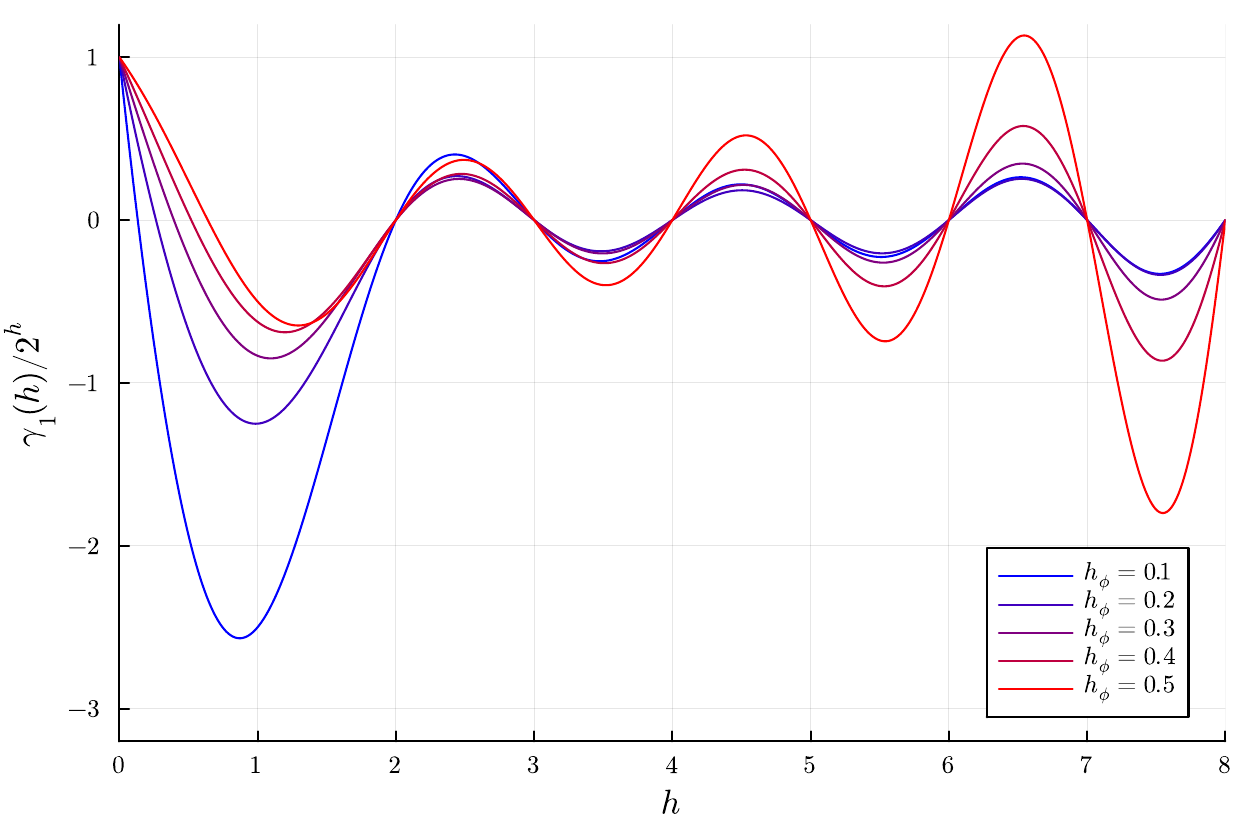}
    \caption{The functional $\gamma_1$ for different $h_\phi$,
    showing how the value at $h = 1$ varies with $h_\phi$.
    The inverse of this value determines the OPE coefficient of an operator with $h = 1$, assuming that all operators in the OPE have integer conformal weight.
    We observe $\gamma_1(1) = (2 h_\phi)^{-1}$ in this case (note the scaling factor $2^{-h}$ in the vertical axis).}
    \label{fig:gamma1}
\end{figure}

The ratio of $\gamma_n(0)$ and $\gamma_n(n)$ depends, of course, on the conformal weight $h_\phi$ of the operators whose OPE is tested. This is shown in Figure~\ref{fig:gamma1}, for $n = 1$. 
In this case, as in Figure~\ref{fig:gammafunctionals} above, we have chosen to normalize the functionals differently so that $\gamma_n(0) = 1$, but the relevant ratio remains the same.
The high numerical precision of our results with $\Lambda = 25$ implies that we can make several empirical observations: 
\begin{itemize}
    \item At $n = 1$, as seen in Figure~\ref{fig:gamma1}, we find
    \begin{equation}
        C_1^2 = 4 h_\phi.
    \end{equation}
    This corresponds however to a spin-one operator that must be discarded from the $\phi \times \phi$ OPE, as discussed above.

    \item At $n = 2$, we have
    \begin{equation}
        C_2^2 = 4 h_\phi^2.
    \end{equation}
    The operator with $h = 2$ is the holomorphic component of the energy-momentum tensor.
    Since it is a conserved current, its 3-point function is determined by a Ward identity. With our conventions specified in Section~\ref{sec:blocks-discontinuity}, this gives $C_2 = 2h_\phi$, in agreement with the observation.

    \item The case $n = 3$ is again a fictitious operator that must be discarded. We observe nevertheless the simple relation
    \begin{equation}
        C_3^2 = \frac{8}{3} h_\phi^3.
    \end{equation}

    \item At $n = 4$, we find
    \begin{equation}
        C_4^2 = \frac{4}{3} h_\phi^4 + \frac{1}{15} h_\phi^2.
    \end{equation}
\end{itemize}
The simple polynomial structure visible in all these cases seems to disappear when $n > 4$. At large $n$, we see that the OPE coefficients approach the generalized free field theory result of eq.~\eqref{eq:GFFcoefficients}. In the limit $h_\phi \to \frac{1}{2}$, all $C_n$ are actually consistent with the free fermion coefficients given by
\begin{equation}
     C_n^2 = \frac{2\Gamma(n)^2}{\Gamma(2n - 1)}
     \qquad
     \text{for} ~ h_\phi = \frac{1}{2}.
\end{equation}

Besides the Ward identity for the energy-momentum tensor that follows from global conformal symmetry, one can actually make use of local conformal invariance to predict the values of the OPE coefficient for all even-$n$ operators in the vacuum module. 
Computations involving the Virasoro algebra with central charge $c$ give for instance~\cite{Perlmutter:2015iya}
\begin{equation}
    C_4^2 = \frac{4 h_\phi^2 (1 + 5 h_\phi)^2}{5(5c + 22)}.
\end{equation}
Equating this with the empirical observation made above, we obtain a rather surprising relation between $c$ and $h_\phi$,
\begin{equation}
    c = \frac{12 h_\phi (h_\phi + 2)}{1 + 20 h_\phi^2} - 2.
    \label{eq:simplecurrentrelation}
\end{equation}
This relation has been verified against similar prediction from Virasoro symmetry at level 6, 8, and 10, even though the numerical results are not fully conclusive since the precision of the functional decays at large $h$.

As mentioned above, eq.~\eqref{eq:simplecurrentrelation} is only true for $h_\phi < \frac{1}{2}$, as the functionals $\gamma_n$ cease to exist above this value. It must be valid in theories that only feature descendants of the identity operator in the OPE $\phi \times \phi$.
In fact, three known cases satisfy indeed the equation:
\begin{itemize}
    \item the free fermion, or equivalently the energy operator $\varepsilon$ in the Ising model, with $h_\phi = \frac{1}{2}$ and $c = \frac{1}{2}$;

    \item the 4-state Potts model with $h_\phi = \frac{1}{4}$ and $c = 1$;

    \item the triplet model realized with a pair of symplectic fermions, with $h_\phi = 0$ and $c = -2$~\cite{Gaberdiel:1998ps, Hogervorst:2016itc}.
\end{itemize}
Whether these are isolated examples or members of a larger family of theories satisfying the relation \eqref{eq:simplecurrentrelation} is an open question.
An analytic construction of the functionals $\gamma_n$ could possibly answer this question.

%%%%%%%%%%%%%%%%%%%%%%%%%%%%%%%%%%%%%%%%%%%%%%%%%%%

\section{Conclusions}
\label{sec:conclusions}

The present work introduces a number of previously-unknown equations that must presumably be satisfied in two-dimensional conformal field theory, together with several non-trivial consistency checks.
Since a summary of results was already given in the introduction, there is no point in repeating it here. We do not want to speculate either on the potential usefulness of these new crossing equations, but highlight instead some potential future developments.

First of all, note that a previously open question has been partially answered in this work, actually by accident: how are the momentum-space conformal blocks in Minkowksi space-time related to the Euclidean position-space blocks. The relation is in fact only established in a special situation with two light-like momenta, but there are good chances that it can be generalized to arbitrary momentum-space kinematics.

Second, the momentum-space crossing equation seems to be particularly well-suited to study generalized free field theory. A natural question is whether it can be used to bootstrap theories in a neighborhood of GFF, such as weakly-relevant flows, in a perturbative expansion~\cite{Zamolodchikov:1987ti, Cappelli:1990, Komargodski:2011xv, Karateev:2024skm}.

Finally, we want to emphasize that none of the methods developed in this work are specific to two-dimensional theories. The analyticity condition originating from causality in Minkowski space-time is valid in any $d \geq 2$, and momentum-space conformal blocks for scalar operators are known in closed-form in any $d > 2$~\cite{Gillioz:2020wgw}. Constructing the momentum-space crossing equation in this case is an interesting problem left for future work.

\subsection*{Acknowledgments}

This paper is the outcome of a long (and no yet finished) project that started exactly 10 years ago under the leadership of Markus Luty. I am thankful to him, as well as to Riccardo Rattazzi and Marco Serone for giving me the opportunity to work on this fascinating subject, even if I could not always honor their patience. I would like to thank also Denis Karateev and Bernardo Zan for helpful discussions, Jo\~ao Penedones and his group at EPFL for their hospitality, and Dalimil Maz\'a{\v c} for pointing to the connection with the contour integrals in cross-ratio space.

\appendix
\section{Discontinuity of the momentum-space integral}
\label{sec:discontinuity}

This appendix details the computation of the discontinuity of the integral
\begin{equation}
    I(\delta) \equiv \int dt
    \left[ e^{i \pi \alpha^+} (t + \delta + i \varepsilon)^{2\beta^+ - 1} + \text{c.c.} \right]
    \left[ e^{i \pi \alpha^-} (t - \delta + i \varepsilon)^{2\beta^- - 1} + \text{c.c.} \right],
\end{equation}
where $\delta$ is a real variable, $\alpha^\pm$ and $\beta^\pm$ real parameters, 
c.c.~denotes the complex conjugate of the previous term,
and the limit $\varepsilon \to 0_+$ is understood. 

The integrand has logarithmic singularities just above and below the contour of integration along the real axis at $\pm \delta$. Expanding the product of the sums in square brackets, there are four terms with two singularities each.
When $\delta \neq 0$, the integral defines an analytic function since the contour can be deformed away from the singularities of the integrand in each of the four terms. When $\delta = 0$, however, the contour is pinched between singularities above and below the real axis. This happens for the two cross-terms, and we find
\begin{equation}
    I(\delta) = e^{i \pi (\alpha^- - \alpha^+)}
    \int dt (t + \delta - i \varepsilon)^{2\beta^+ - 1}
    (t - \delta + i \varepsilon)^{2\beta^- - 1}
    + \text{c.c.} + \text{analytic in}~\delta.
\end{equation}
The discontinuity is then defined as the difference 
\begin{equation}
    \disc I(\delta) \sim I(\delta) - I(-\delta),
    \qquad\qquad
    \text{as} ~ \delta \to 0_+,
\end{equation}
where the limit must be understood in the asymptotic sense.
Using reflection $t \to -t$, one can see that
\begin{equation}
    I(-\delta) = e^{i \pi (\alpha^+ - \alpha^-)} e^{2i \pi (\beta^+ - \beta^-)}
    \int dt (t + \delta - i \varepsilon)^{2\beta^+ - 1}
    (t - \delta + i \varepsilon)^{2\beta^- - 1}
    + \text{c.c.} + \text{analytic in}~\delta.
\end{equation}
and thus
\begin{equation}
    \begin{split}
        \disc I(\delta)
        &= 2 \sin \left[ \pi (\alpha^- - \alpha^+ + \beta^- - \beta^+) \right]
        \\
        & \quad \times
        \left[ e^{i \pi (2\beta^+ - 2\beta^- + 1) / 2}
        \int dt (t + \delta - i \varepsilon)^{2\beta^+ - 1}
        (t - \delta + i \varepsilon)^{2\beta^- - 1}
        + \text{c.c.} \right]
    \end{split}
\end{equation}
The integral can be solved by rotating the contour of integration towards the imaginary axis,
\begin{equation}
    \begin{split}
        \disc I(\delta)
        &= 4 \sin \left[ \pi (\alpha^- - \alpha^+ + \beta^- - \beta^+) \right]
        \sin \left[ \pi \left( \tfrac{1}{2} - \beta^+ - \beta^- \right) \right]
        \delta^{2\beta^+ + 2\beta^- - 1}
        \\
        & \quad\times
        \int dt (1 + i t)^{2\beta^+ - 1} (1 - i t)^{2\beta^- - 1}.
    \end{split}
\end{equation}
The remaining integral is a well-known representation of the $\beta$ function, hence
\begin{equation}
    \begin{split}
        \disc I(\delta)
        &= 8 \sin \left[ \pi (\alpha^- - \alpha^+ + \beta^- - \beta^+) \right]
        \sin \left[ \pi \left( \tfrac{1}{2} - \beta^+ - \beta^- \right) \right]
        \\
        & \quad \times
        \frac{\Gamma(1 - 2\beta^+ - 2\beta^-)}{\Gamma(1 - 2\beta^+) \Gamma(1 - 2\beta^-)}
        (2\delta)^{2\beta^+ + 2\beta^- - 1}
    \end{split}
\end{equation}
which can be rewritten as
\begin{equation}
    \disc I(\delta)
    = \frac{4\pi \sin \left[ \pi (\alpha^- - \alpha^+ + \beta^- - \beta^+) \right]
    (2\delta)^{2\beta^+ + 2\beta^- - 1}}
    {\sin\left[ \pi (\beta^+ + \beta^-) \right]
    \Gamma(2\beta^+ + 2\beta^-) \Gamma(1 - 2\beta^+) \Gamma(1 - 2\beta^-)}.
\end{equation}

\section{Hypergeometric functions and Jacobi polynomials}
\label{sec:integrals}

In this appendix, we compute the integral
\begin{equation}
    \mathcal{I} = \frac{j!}{\Gamma(\beta + j)}
    \int_0^1 dw \, w^{\gamma - 1} (1-w)^{\beta - 1} 
    P_j^{(\alpha - 1, \beta - 1)}(1 - 2w)
    {}_2F_1(a, b; c; w),
    \label{eq:I:def}
\end{equation}
with $\alpha, \beta, \gamma > 0$,
and show that it is equal to a generalized hypergeometric function evaluated at arguement one,
\begin{equation}
    \mathcal{I} = \frac{(\alpha - \gamma)_j \Gamma(\gamma)}
    {\Gamma(\beta + \gamma + j)}
    {}_4F_3\left( \begin{array}{c}
        a, b, \gamma, \gamma - \alpha + 1 \\
        c, \beta + \gamma + j, \gamma - \alpha - j + 1
    \end{array}; 1 \right).
    \label{eq:I:result}
\end{equation}
In the special case $\alpha = \gamma = c$, we also show that this is equal to
\begin{equation}
    \mathcal{I} = \frac{(-1)^j (a)_j (b)_j \Gamma(\alpha) \Gamma(\alpha + \beta - a - b)}
    {\Gamma(\alpha + \beta - a + j) \Gamma(\alpha + \beta - b + j)}.
    \label{eq:I:special}
\end{equation}

\subsubsection*{Proof of the general case}

Any Jacobi polynomial can be expanded in a finite sum as
\begin{equation}
    P_j^{(\alpha - 1, \beta - 1)}(1 - 2w) = \sum_{i = 0}^j
    \frac{(-1)^i (\alpha + i)_{j-i} (\beta + j - i)_i}{i! (j-i)!} w^i (1-w)^{j-i}.
\end{equation}
Together with the definition of the hypergeometric series for ${}_2F_1(a, b; c; w)$, this can be used to write
\begin{equation}
    \mathcal{I} = \frac{j!}{\Gamma(\beta + j)} \sum_{k=0}^\infty
    \frac{(a)_k (b)_k}{k! (c)_k}
    \sum_{i = 0}^j
    \frac{(-1)^i (\alpha + i)_{j-i} (\beta + j - i)_i}{i! (j-i)!} 
    \int_0^1 dw \,
    w^{\gamma + i + k - 1} (1-w)^{\beta + j - i - 1}.
\end{equation}
We have permuted the infinite sum and the integral since the latter converges term-by-term: we work under the assumptions $\beta, \gamma > 0$, and $i \leq j$, and therefore
\begin{equation}
    \int_0^1 dw \,
    w^{\gamma + i + k - 1} (1-w)^{\beta + j - i - 1}
    = \frac{\Gamma(\gamma + i + k) \Gamma(\beta + j - i)}
    {\Gamma(\beta + \gamma + j + k)}.
\end{equation}
After some simplifications using $\Gamma(x + n) = (x)_n \Gamma(x)$, we arrive at
\begin{equation}
    \mathcal{I} = \frac{j! \Gamma(\gamma)}{\Gamma(\beta + \gamma + j)}
    \sum_{k=0}^\infty
    \frac{(a)_k (b)_k (\gamma)_k }{k! (c)_k (\beta + \gamma + j)_k}
    \sum_{i = 0}^j
    \frac{(-1)^i (\alpha + i)_{j-i} (\gamma + k)_i}{i! (j-i)!}.
    \label{eq:I:doublesum}
\end{equation}
The sum over $i$ is actually simple: using the identity
\begin{equation}
    (x + y)_j = \sum_{i = 0}^j \frac{(x)_i (y)_{j-i}}{i! (j - i)!},
\end{equation}
we find
\begin{equation}
    \begin{split}
        (\alpha - \gamma - k)_j &= (-1)^j (1 - \alpha + \gamma + k - j)_j
        \\
        &= (-1)^j \sum_{i = 0}^j \frac{(1 - \alpha - j)_{j-i} (\gamma + k)_i }{i! (j - i)!}
        = \sum_{i = 0}^j \frac{(-1)^i (\alpha + i)_{j-i} (\gamma + k)_i }{i! (j - i)!}.
    \end{split}
\end{equation}
The right-hand side is precisely the sum over $i$ in eq.~\eqref{eq:I:doublesum}, so that
\begin{equation}
    \mathcal{I} = \frac{\Gamma(\gamma)}{\Gamma(\beta + \gamma + j)}
    \sum_{k=0}^\infty
    \frac{(a)_k (b)_k (\gamma)_k (\alpha - \gamma - k)_j}{k! (c)_k (\beta + \gamma + j)_k}.
    \label{eq:I:singlesum}
\end{equation}
Using a few more identities of the Pochhammer symbol, we can show that 
\begin{equation}
    (\alpha - \gamma - k)_j
    = \frac{(\alpha - \gamma)_j (\gamma - \alpha + 1)_k}{(\gamma - \alpha - j + 1)_k}
\end{equation}
and thus
\begin{equation}
    \mathcal{I} = \frac{(\alpha - \gamma)_j \Gamma(\gamma)}
    {\Gamma(\beta + \gamma + j)}
    \sum_{k=0}^\infty
    \frac{(a)_k (b)_k (\gamma)_k (\gamma - \alpha + 1)_k}
    {k! (c)_k (\beta + \gamma + j)_k (\gamma - \alpha - j + 1)_k}.
\end{equation}
This is the definition of the ${}_4F_3$ hypergeometric series appearing in eq.~\eqref{eq:I:result}, which concludes the proof.

\subsubsection*{Proof of the special case}

The case $\alpha = \gamma = c$ is not straightforward to evaluate. Already when $\alpha = \gamma$, the ${}_4F_3$ hypergeometric is ill-defined for all $j > 0$. In this case, we go back to eq.~\eqref{eq:I:singlesum}, which becomes
\begin{equation}
    \mathcal{I} = \frac{\Gamma(\alpha)}{\Gamma(\alpha + \beta + j)}
    \sum_{k=0}^\infty
    \frac{(a)_k (b)_k (\alpha)_k (- k)_j}{k! (c)_k (\alpha + \beta + j)_k},
\end{equation}
and rewrite it as
\begin{equation}
    \mathcal{I} = \frac{(-1)^j \Gamma(\alpha)}{\Gamma(\alpha + \beta + j)}
    \sum_{k=0}^\infty
    \frac{(a)_k (b)_k (\alpha)_k}{(k-j)! (c)_k (\alpha + \beta + j)_k}.
\end{equation}
All terms with $k < j$ vanish, so we can shift the sum by $j$ and obtain
\begin{equation}
    \mathcal{I} = 
    \frac{(-1)^j (a)_j (b)_j \Gamma(\alpha + j)}{(c)_j \Gamma(\alpha + \beta + 2j)}
    \sum_{k=0}^\infty
    \frac{(a + j)_k (b + j)_k (\alpha + j)_k}{k! (c + j)_k (\alpha + \beta + 2j)_k},
\end{equation}
in which we recognize a hypergeometric series
\begin{equation}
    \mathcal{I} = \frac{(-1)^j (a)_j (b)_j \Gamma(\alpha + j)}
    {(c)_j \Gamma(\alpha + \beta + 2j)}
    {}_3F_2\left( \begin{array}{c}
        a + j, b + j, \alpha + j \\
        c + j, \alpha + \beta + 2j
    \end{array}; 1 \right).
\end{equation}
Now when $c = \alpha$, this is
\begin{equation}
    \mathcal{I} = \frac{(-1)^j (a)_j (b)_j \Gamma(\alpha)}
    {\Gamma(\alpha + \beta + 2j)}
    {}_2F_1(a + j, b + j; \alpha + \beta + 2j; 1),
\end{equation}
and eq.~\eqref{eq:I:special} is recovered using Gauss' formula, valid when $\alpha + \beta > a + b$.
If the latter condition is not satisfied, we can proceed by analytic continuation in $a$ and $b$, starting from $a = b = 0$. Both the integral \eqref{eq:I:def} and the result \eqref{eq:I:special} are analytic in $a, b \geq 0$ away from a few singular points, so we conclude that the two are equal.

Alternatively, we can obtain the same result from eq.~\eqref{eq:I:result}
using some more involved hypergeometric identities. When $c = \gamma$, we have
\begin{equation}
    \mathcal{I} = \frac{(\alpha - \gamma)_j \Gamma(\gamma)}
    {\Gamma(\beta + \gamma + j)}
    {}_3F_2\left( \begin{array}{c}
        a, b, \gamma - \alpha + 1 \\
        \beta + \gamma + j, \gamma - \alpha - j + 1
    \end{array}; 1 \right).
\end{equation}
We can then use an identity of the ${}_3F_2$ hypergeometric known as Thomae's theorem~\cite{Bailey:1935} to rewrite this as 
\begin{equation}
    \mathcal{I} = \frac{(-1)^j \Gamma(\beta + \gamma - a - b) \Gamma(\gamma)}
    {\Gamma(\beta + \gamma - a) \Gamma(\beta + \gamma - b)}
    {}_3F_2\left( \begin{array}{c}
        -j, \alpha + \beta + j - 1, \beta + \gamma - a - b \\
        \beta + \gamma - a, \beta + \gamma - b
    \end{array}; 1 \right).
\end{equation}
This new form is well-defined when $\gamma = \alpha$, giving
\begin{equation}
    \mathcal{I} = \frac{(-1)^j \Gamma(\alpha) \Gamma(\alpha + \beta - a - b)}
    {\Gamma(\alpha + \beta - a) \Gamma(\alpha + \beta - b)}
    {}_3F_2\left( \begin{array}{c}
        -j, \alpha + \beta + j - 1, \alpha + \beta - a - b \\
        \alpha + \beta - a, \alpha + \beta - b
    \end{array}; 1 \right).
\end{equation}
Saalschütz's theorem~\cite{Bailey:1935} finally says that 
\begin{equation}
    {}_3F_2\left( \begin{array}{c}
        -j, \alpha + \beta + j - 1, \alpha + \beta - a - b \\
        \alpha + \beta - a, \alpha + \beta - b
    \end{array}; 1 \right) = 
    \frac{(a)_j (b)_j}{(\alpha + \beta - a)_j (\alpha + \beta - b)_j},
\end{equation}
and we recover the result of eq.~\eqref{eq:I:special}.

\section{Euclidean connection with distinct operators}
\label{sec:Euclideanconnection:distinct}

This appendix shows how the momentum-space crossing equation~\eqref{eq:2dcrossing} with the conformal blocks given by eqs.~\eqref{eq:S} and \eqref{eq:T} can be obtained from the position-space crossing equation.
Since the complete form of the crossing equation with four distinct operators is not easily found in the literature, we begin with a short review of its derivation.

\subsubsection*{Crossing symmetry on a line with distinct operators}

Consider a correlation function with holomorphic operators $\chi_i$ inserted at points $z_i$ on a line, with operators 2 and 3 surrounded by 1 and 4,
\begin{equation}
    z_1 > z_2 > z_4,
    \qquad
    z_1 > z_3 > z_4.
\end{equation}
We choose to parameterize the 4-point function as
\begin{equation}
    \langle 0 | \chi_1(z_1) \chi_2(z_2) \chi_3(z_3) \chi_4(z_4) | 0 \rangle
    = \frac{(z_1 - z_2)^{h_3 + h_4 - h_1 - h_2} (z_1 - z_3)^{h_2 + h_4 - h_1 - h_3}}
    {(z_1 - z_4)^{2h_4} (z_2 - z_3)^{h_2 + h_3 + h_4 - h_1}} G(z),
    \label{eq:Euclidean4pt:symmetric}
\end{equation}
where $z$ is the conformal cross-ratio
\begin{equation}
    z = \frac{(z_1 - z_2)(z_3 - z_4)}{(z_1 - z_4) (z_3 - z_2)}
    \qquad\Leftrightarrow\qquad
    1 - z = \frac{(z_1 - z_3)(z_2 - z_4)}{(z_1 - z_4) (z_2 - z_3)}.
\end{equation}
Note that $z \leftrightarrow 1-z$ under the exchange $z_2 \leftrightarrow z_3$.

Configurations of points with $z_1 > z_2 > z_3 > z_4$ correspond to $G(z)$ in the interval $z \in (-\infty, 0)$. However, since $z_3 - z_4$ does not appear explicitly in eq.~\eqref{eq:Euclidean4pt:symmetric}, we can extend this definition by analytic continuation in $z$ to a configuration of the form $z_1 > z_2 > z_4 > z_3$, which corresponds to $z \in (0, 1)$.
In this domain, we can use the conformal block expansion in the channel $(12)(34)$.
Referring to the seminal work of Dolan and Osborn~\cite{Dolan:2000ut, Dolan:2003hv}, the expansion is usually given using the parameterization
\begin{equation}
    \begin{split}
        \langle \chi_1(z_1) \chi_2(z_2) \chi_4(z_4) \chi_3(z_3)  \rangle
        &= \frac{1}{(z_1 - z_2)^{h_1 + h_2} (z_4 - z_3)^{h_3 + h_4}}
        \\
        & \quad \times
        \left( \frac{z_1 - z_3}{z_2 - z_3} \right)^{h_2 - h_1}
        \left( \frac{z_1 - z_4}{z_1 - z_3} \right)^{h_3 - h_4} G_{DO}(z),
        \label{eq:Euclidean4pt:s}
    \end{split}
\end{equation}
which is related to ours by
\begin{equation}
    G_{DO}(z) = z^{h_3 + h_4} G(z).
\end{equation}
The expansion of $G(z)$ is therefore given by
\begin{equation}
    G(z) = \sum_\O C_{12\O} C_{34\O}^* g_h^{(s)}(z)
    \qquad
    \text{for} ~ z_2 > z_3,
    \label{eq:EuclideanOPE:s}
\end{equation}
where the s-channel conformal blocks are
\begin{equation}
    g_h^{(s)}(z) = z^{h - h_3 - h_4} {}_2F_1(h + h_2 - h_1, h + h_4 - h_3; 2h; z).
    \label{eq:Euclideanblock:s}
\end{equation}

A similar argument can be made in the channel $(13)(24)$, starting from the configuration $z_1 > z_3 > z_2 > z_4$, where $z \in (1, \infty)$: since eq.~\eqref{eq:Euclidean4pt:symmetric} does not depend explicitly on $z_2 - z_4$, the configuration $z_1 > z_3 > z_4 > z_2$ corresponding again to $z \in (0,1)$ can be reached by analytic continuation in $z$. In this case, we can use the conformal block expansion
\begin{equation}
    G(1 - z) = \sum_\O C_{13\O} C_{24\O}^* g_h^{(t)}(1 - z),
    \qquad
    \text{for} ~ z_2 < z_3,
    \label{eq:EuclideanOPE:t}
\end{equation}
with
\begin{equation}
    g_h^{(t)}(z) = z^{h - h_2 - h_4} {}_2F_1(h + h_3 - h_1, h + h_4 - h_2; 2h; z).
\end{equation}
Note that $G(z)$ and $G(1-z)$ defined respectively in eqs.~\eqref{eq:EuclideanOPE:s} and \eqref{eq:EuclideanOPE:t} are not equal at this stage: they differ by a phase related to the exchange of $z_2$ and $z_3$ in the denominator of eq.~\eqref{eq:Euclidean4pt:symmetric}.

\subsubsection*{Crossing symmetry in 2d with distinct operators}

The two-dimensional case with generic, non-holomorphic operators $\O_i$ involves another copy of the correlation function expressed in terms of the coordinates $\zbar_i$ and conformal weights $\hbar_i$. The same argument as above can be repeated in the anti-holomorphic sector since $z$ and $\zbar$ are independent variables. 
However, we must consider the reverse ordering of points to ensure that they remain space-like separated throughout the argument, namely
\begin{equation}
    \zbar_1 < \zbar_2 < \zbar_4,
    \qquad
    \zbar_1 < \zbar_3 < \zbar_4.
\end{equation}
This amounts to inverting $1 \leftrightarrow 4$ and $2 \leftrightarrow 3$ in the anti-holomorphic conformal blocks, so that
\begin{align}
    g_\hbar^{(s)}(\zbar) &= \zbar^{\hbar - \hbar_1 - \hbar_2}
    {}_2F_1(\hbar + \hbar_1 - \hbar_2, \hbar + \hbar_3 - \hbar_4; 2\hbar; \zbar),
    \\
    g_\hbar^{(t)}(\zbar) &= \zbar^{\hbar - \hbar_1 - \hbar_3}
    {}_2F_1(\hbar + \hbar_1 - \hbar_3, \hbar + \hbar_2 - \hbar_4; 2\hbar; \zbar).
\end{align}
Combining the two sides, we obtain finally the crossing equation
\begin{equation}
    \sum_\O C_{12\O} C_{34\O}^* g_h^{(s)}(z) g_\hbar^{(s)}(\zbar)
    = \sum_\O C_{13\O} C_{24\O}^* g_h^{(t)}(1 - z) g_\hbar^{(t)}(1 - \zbar).
    \label{eq:crossing:z:distinct}
\end{equation}
Note that the phase associated with the exchanges $z_2 \leftrightarrow z_3$ in the denominator of eq.~\eqref{eq:Euclidean4pt:symmetric} combines with its anti-holomorphic counterpart under $\zbar_2 \leftrightarrow \zbar_3$ to form a factor of $(-1)^{\ell_1 + \ell_2 + \ell_3 + \ell_4}$. Since the total spin must be even, this is equal to $+1$. This is why the two expansion must agree on the interval $z, \zbar \in (0,1)$, and by analytic continuation in the entire complex planes apart from the branch cuts at $z, \zbar \in (-\infty, 0) \cup (1, \infty)$.
In the case of identical external operators, we recover the crossing equation \eqref{eq:crossing:z} with conformal blocks that are identical in the s- and t-channel, both corresponding to eq.~\eqref{eq:g}.

\subsubsection*{Discontinuities at the branch cuts}

The next step is to evaluate the discontinuity of the conformal block along the branch cut $z \in (1, \infty)$, on both sides of the crossing equation. Taking $z = 1/(1-x) + i \varepsilon$ with $x \in (0,1)$, we define
\begin{align}
    f_h^{(s)}(x) \equiv \lim_{\varepsilon \to 0_+} \frac{1}{\pi}
    \im g_h^{(s)}\left( \frac{1}{1 - x} + i \varepsilon \right),
    \qquad
    f_h^{(t)}(x) \equiv \lim_{\varepsilon \to 0_+} \frac{1}{\pi}
    \im g_h^{(t)}\left( \frac{x}{x-1} - i \varepsilon \right).
\end{align}
For the t-channel block, this is simply the discontinuity associated with the non-integer power of $(1-z)$, and we obtain directly
\begin{equation}
    f_h^{(t)}(x) = \frac{\sin[\pi (h_2 + h_4 - h)]}{\pi}
    \left( \frac{x}{1-x} \right)^{h - h_2 - h_4}
    {}_2F_1\left( h + h_3 - h_1, h + h_4 - h_2; 2h; \frac{x}{x - 1} \right),
\end{equation}
which is equivalent to
\begin{equation}
    f_h^{(t)}(x) = \frac{\sin[\pi (h_2 + h_4 - h)]}{\pi}
    x^{h - h_2 - h_4} (1-x)^{2h_4}
    {}_2F_1( h + h_1 - h_3, h + h_4 - h_2; 2h; x).
\end{equation}
For the s-channel block, we first use an identity of the hypergeometric function to write
\begin{equation}
    \begin{split}
        g_h^{(s)}\left( \frac{1}{1-x} + i \varepsilon \right) &= 
        \frac{\Gamma(2h) \Gamma(h_1 + h_3 - h_2 - h_4)}{\Gamma(h + h_1 - h_2) \Gamma(h + h_3 - h_4)}
        (1-x)^{h_2 + h_3 + h_4 - h_1}
        \\
        & \quad\qquad\times
        {}_2F_1(h + h_2 - h_1, 1 - h + h_2 - h_1; h_2 + h_4 - h_1 - h_3 + 1; x)
        \\
        & \quad
        + \frac{\Gamma(2h) \Gamma(h_2 + h_4 - h_1 - h_3)}{\Gamma(h + h_2 - h_1) \Gamma(h + h_4 - h_3)}
        (-x - i \varepsilon)^{h_1 + h_3 - h_2 - h_4}
        (1 - x)^{2h_4}
        \\
        & \quad\qquad\times
        {}_2F_1(h + h_1 - h_2, 1 - h + h_1 - h_2; h_1 + h_3 - h_2 - h_4 + 1; x).
    \end{split}
\end{equation}
Of the two terms, only the second one displays a discontinuity, and we obtain finally
\begin{equation}
    \begin{split}
        f_h^{(s)}(x) &= 
        \frac{\Gamma(2h) x^{h_1 + h_3 - h_2 - h_4} (1 - x)^{2h_4}}
        {\Gamma(h + h_2 - h_1) \Gamma(h + h_4 - h_3)
        \Gamma(h_1 + h_3 - h_2 - h_4 + 1)}
        \\
        & \quad\times
        {}_2F_1(h + h_1 - h_2, 1 - h + h_1 - h_2; h_1 + h_3 - h_2 - h_4 + 1; x).
    \end{split}
\end{equation}

\subsubsection*{Analytic functionals}

In the case of identical operators, the functionals of Ref.~\cite{Mazac:2016qev} were found integrating against Legendre polynomials. In this case, we must use a more general Jacobi polynomial and define
\begin{equation}
    \omega_j[f] = \frac{(-1)^j j!}{\Gamma(1 + \beta + j)}
    \int\limits_0^1 dx 
    (1-x)^{h_1 - h_2 - h_3 - h_4} P_j^{(\alpha, \beta)}(1 - 2x) f(x)
\end{equation}
where
\begin{equation}
    \alpha = h_1 + h_3 - h_2 - h_4,
    \qquad\qquad
    \beta = h_1 + h_4 - h_2 - h_3.
\end{equation}
The reason for this choice is that $f_h^{(s)}$ is itself proportional to a Jacobi polynomial when $h = h_1 - h_2 + 1 + n$:
\begin{equation}
    (1-x)^{h_1 - h_2 - h_3 - h_4} f_h^{(s)}(x)
    = \frac{\Gamma(2h)}{\Gamma(1 + \alpha + n) \Gamma(1 + \beta + n)}
    x^\alpha (1-x)^\beta
    P_n^{(\alpha, \beta)}(1 - 2x).
\end{equation}
Note that $\alpha = \beta = 0$ when the four operators are identical, and the Jacobi polynomial $P_n^{(0,0)}$ is simply a Legendre polynomial in this case.

Using the results of Appendix~\ref{sec:integrals}, we find the equivalence
\begin{equation}
    \omega_j[f_h^{(s)}] = \widetilde{S}_{h, j},
    \qquad\qquad
    \omega_j[f_h^{(t)}] = \widetilde{T}_{h, j},
\end{equation}
where $\widetilde{S}_{h, j}$ and $\widetilde{T}_{h, j}$ are the functionals given in eqs.~\eqref{eq:Sjtilde} and \eqref{eq:Tjtilde}.
There are similar results for the anti-holomorphic blocks, upon replacing $h_i \to \hbar_{5-i}$.
This concludes the proof that the momentum-space conformal blocks are related to a particular integral transform of the position-space blocks along the discontinuity of the Euclidean sheet.

\section{A special case: the 4-state Potts model}
\label{sec:4statePotts}

An interesting observation can be made inspecting the conformal blocks \eqref{eq:S:identical} and \eqref{eq:T:identical} carefully: when $h_\phi =\frac{1}{4}$, a peculiar identity of the hypergeometric function lets us rewrite them using trigonometric functions. After making the change of variable $w = (\sin \theta)^2$ with $\theta \in \left(0, \frac{\pi}{2} \right)$, we have
\begin{align}
    S_h\big( (\sin\theta)^2 \big) &= \frac{2}{\sqrt{\pi}} \frac{\Gamma(2h)}{\Gamma(h)^2}
    \frac{\cos\left[ (2h-1) \theta \right]}{\sin(2\theta)},
    \\
    T_h\big( (\sin\theta)^2 \big) &= \frac{2}{\sqrt{\pi}} \frac{\Gamma(2h)}{\Gamma(h)^2}
    \frac{\cos(\pi h) \left( \tan \frac{\theta}{2} \right)^{2h-1}}{\sin(2\theta)}.
\end{align}
In this case, the Jacobi polynomials entering the functional \eqref{eq:functional:omegahat} are proportional to Chebyshev polynomials, which are simply cosines of even multiples of $\theta$,
\begin{equation}
    \widehat{P}_j\left( (\sin \theta)^2 \right)
    = \frac{(2j)!}{2^{2j} (j!)^2} \cos(2 j \theta)
\end{equation}
so that
\begin{equation}
    \widehat{\omega}_j[f] = \frac{1}{\sqrt{\pi}}
    \int_0^{\pi/2} d\theta \sin(2\theta) \cos(2 j \theta)
    f\left( (\sin\theta)^2 \right).
    \label{eq:functional:omegatilde:4statePotts}
\end{equation}
This situation is realized in the 4-state Potts models~\cite{Gorbenko:2018ncu, Gorbenko:2018dtm}: the energy operator $\varepsilon$ has $(h, \hbar) = \left( \frac{1}{4}, \frac{1}{4} \right)$, and its OPE only contains the vacuum module, namely operators with integer conformal weights.
This means that the OPE coefficients of the theory solve the equation
\begin{equation}
    \sum_{n = 0}^\infty c_n \left\{ \cos\left[ (2n-1) \theta \right]
    - (-1)^n \left( \tan \tfrac{\theta}{2} \right)^{2n-1} \right\} = 0.
\end{equation}
In spite of its relative simplicity, we were not able to solve this equation analytically. A numerical solution is provided by the functionals $\gamma_n$ of Section~\ref{sec:bootstrapfunctionals}.

\subsubsection*{Analyticity in $j$}

The simple form of the functional $\widehat{\omega}_j$ in eq.~\eqref{eq:functional:omegatilde:4statePotts} lets us compute $S_{h, j}$ and $T_{h,j}$ at arbitrary $j$, not necessarily integer.
We find
\begin{equation}
    S_{h, j} = \frac{\Gamma(2h)}{\Gamma(h)^2}
    \frac{\left(h - \frac{1}{2} \right) \cos(\pi j) \cos(\pi h)
    + j \sin(\pi j) \sin(\pi h)}
    {\pi \left(h - j - \frac{1}{2} \right) \left(h + j - \frac{1}{2} \right)},
\end{equation}
which, at half-integer $j$, vanishes for all integer $h$ except $h = j + \frac{1}{2}$,
as well as
\begin{equation}
    \begin{split}
        T_{h, j} = \frac{\Gamma(2h)}{\Gamma(h)^2} \frac{\cos(\pi h)}{\sin(\pi h)}
        \Big[ & {}_4F_3\left( -j, j, h, 1-h; \tfrac{1}{2}, \tfrac{1}{2}, 1; 1\right)
        \\
        & - (2h-1) {}_4F_3\left( \tfrac{1}{2} - j, \tfrac{1}{2} + j, h, 1-h;
        \tfrac{1}{2}, 1, \tfrac{3}{2}; 1\right) \Big].
    \end{split}
\end{equation}
In this form, $T_{h,j}$ has an apparent singularity at integer $h$ due to the sine in the denominator.
However, the numerator vanishes precisely at integer values of $h$: the terms inside the square brackets are two equivalent representations of a Wilson polynomial of degree $h-1$ in $j$, and the limit of integer $h$ is therefore well-defined. Away from integer $h$, the two ${}_4F_3$ are distinct functions~\cite{Wilson:1980}.

\bibliographystyle{utphys}
\bibliography{biblio}

\end{document}